\documentclass[aps,prb,twocolumn,superscriptaddress,showpacs,floatfix]{revtex4-2}
\usepackage{comment}
\usepackage{physics}
\usepackage{amsmath,amssymb,amsfonts,bm,color,xcolor,graphicx}
\usepackage{float}
\usepackage{textcase}
\usepackage{dcolumn}% Align table columns on decimal point
\UseRawInputEncoding 
\usepackage{dsfont}
\usepackage{blkarray}
\usepackage{hhline}                           
\usepackage[caption=false]{subfig}
\usepackage{multirow}

\newcommand{\be}{\begin{equation}}
\newcommand{\ee}{\end{equation}}
\newcommand{\bk}{{{\bf{k}}}}

\newcommand{\beal}{\begin{align}}
\newcommand{\eeal}{\end{align}}

\newcommand{\upa}{\uparrow}
\newcommand{\dna}{\downarrow}

\newcommand{\dg}{{\dagger}}
\newcommand{\pdg}{{\phantom\dagger}}

\let\Vector\vec
\renewcommand{\vec}[1]{\mathbf{#1}}

\def\ua{\uparrow}
\def\da{\downarrow}

\def\bS{\vec S}

\usepackage{bm}

\usepackage{hyperref}

\newcommand{\beginsupplement}{%
  \setcounter{section}{0}
  \renewcommand{\thesection}{S\arabic{section}}%
  \setcounter{subsection}{0}
  \renewcommand{\thesubsection}{S\arabic{section}.\arabic{subsection}}%
  \setcounter{equation}{0}
  \renewcommand{\theequation}{S\arabic{equation}}%
  \setcounter{figure}{0}
  \renewcommand{\thefigure}{S\arabic{figure}}%
  \setcounter{table}{0}
  \renewcommand{\thetable}{S\arabic{table}}%
}

\begin{document}

\preprint{APS/123-QED}
\title{Two-orbital $e_g$ model with bond-dependent 
spin-orbit coupling: A playground for emergent band topology, Kitaev magnetism, and magnetoelectricity}
\author{YuZheng Xie}
\email{yz.xie@mail.utoronto.ca}
\affiliation{Department of Physics, University of Toronto, 60 St. George Street, Toronto, ON, M5S 1A7 Canada.}
\author{Manoj Gupta}
\email{gpta.mnj@gmail.com}
\affiliation{Department of Condensed Matter Physics and Materials Science, S.N. Bose National Centre for Basic Sciences, Kolkata 700098, India.}
\author{Arun Paramekanti}
\email{arun.paramekanti@utoronto.ca}
\affiliation{Department of Physics, University of Toronto, 60 St. George Street, Toronto, ON, M5S 1A7 Canada.}
\author{Tanusri Saha-Dasgupta}
\email{tanusri@bose.res.in}
\affiliation{Department of Condensed Matter Physics and Materials Science, S.N. Bose National Centre for Basic Sciences, Kolkata 700098, India.}

\date{\today}

\begin{abstract}
Inspired by the electronic structure of compounds like nickel dihalides Ni$X_2$ ($X$=Cl, Br, I), we propose a low-energy two-orbital $e_g$ model featuring bond-dependent spin-orbit terms, driven by atomic spin-orbit coupling on the ligand $X$. We demonstrate that this model hosts a rich array of phenomena. In the non-interacting band limit, spin-orbit-derived spin-dependent and spin-flip hopping terms produce topological bands with spin-Chern numbers $C_s=\pm 2, \pm 4$, and higher order topological states with fractional corner charges, respectively. In the half-filled Mott insulator limit, we recover a spin-$1$ Hamiltonian with bond-dependent Kitaev exchange interactions. We explore the
magnetoelectric effect in this two-orbital model using symmetry-based perspective and microscopic calculations, going beyond
the generalized Katsura-Nagaosa-Balatsky theory for the single-orbital case.
Our work may be relevant to study of doping, strain, or pressure on Ni$X_2$ and related materials.
\end{abstract}

\maketitle

\section{Introduction}

The role of ligands in transition metal complexes has been traditionally assumed to be passive: dictating their chemical stability,
bonding, lattice geometry, and the local electronic structure via crystal-field splitting of the transition
metal $d$-orbitals \cite{Fazekas1999,Khomskii_2014}. This view has been challenged in transition
metal oxides, especially in late transition metal oxides with strong covalency between metal and oxygen. Oxygen, which
is a ligand, is found to play a more active role in terms of the formation of the Zhang-Rice singlet state in 
cuprates \cite{zhang1988effective}, or
driving the site-selective Mott transition through charge disproportionation at the ligand site \cite{nickelate}.

The phenomenology becomes even more intriguing in systems where the ligand site is occupied by a heavy element having large atomic spin-orbit coupling (SOC). Many layered, van der Waals transition metal chalcogenides or halides exhibit non-trivial topological or magnetic properties, driven by $4p$ or $5p$ block chalcogens (Se, Te), such as in WTe$_2$ or WSe$_2$ \cite{ligand1}, or halides (Br, I) as in CrI$_3$ \cite{ligand2}. In this context, the situation for late transition metal complexes in octahedral coordination with ligands is special. Here, the physics at transition metal sites is governed by the crystal-field split $e_g$ orbitals, which have completely quenched angular momentum. Remarkably, strong hybridization of these $e_g$-orbitals with the heavy ligand $p$-orbitals can reactivate spin-orbit phenomena.

\begin{figure}[t]
    \includegraphics[width=0.45\textwidth]{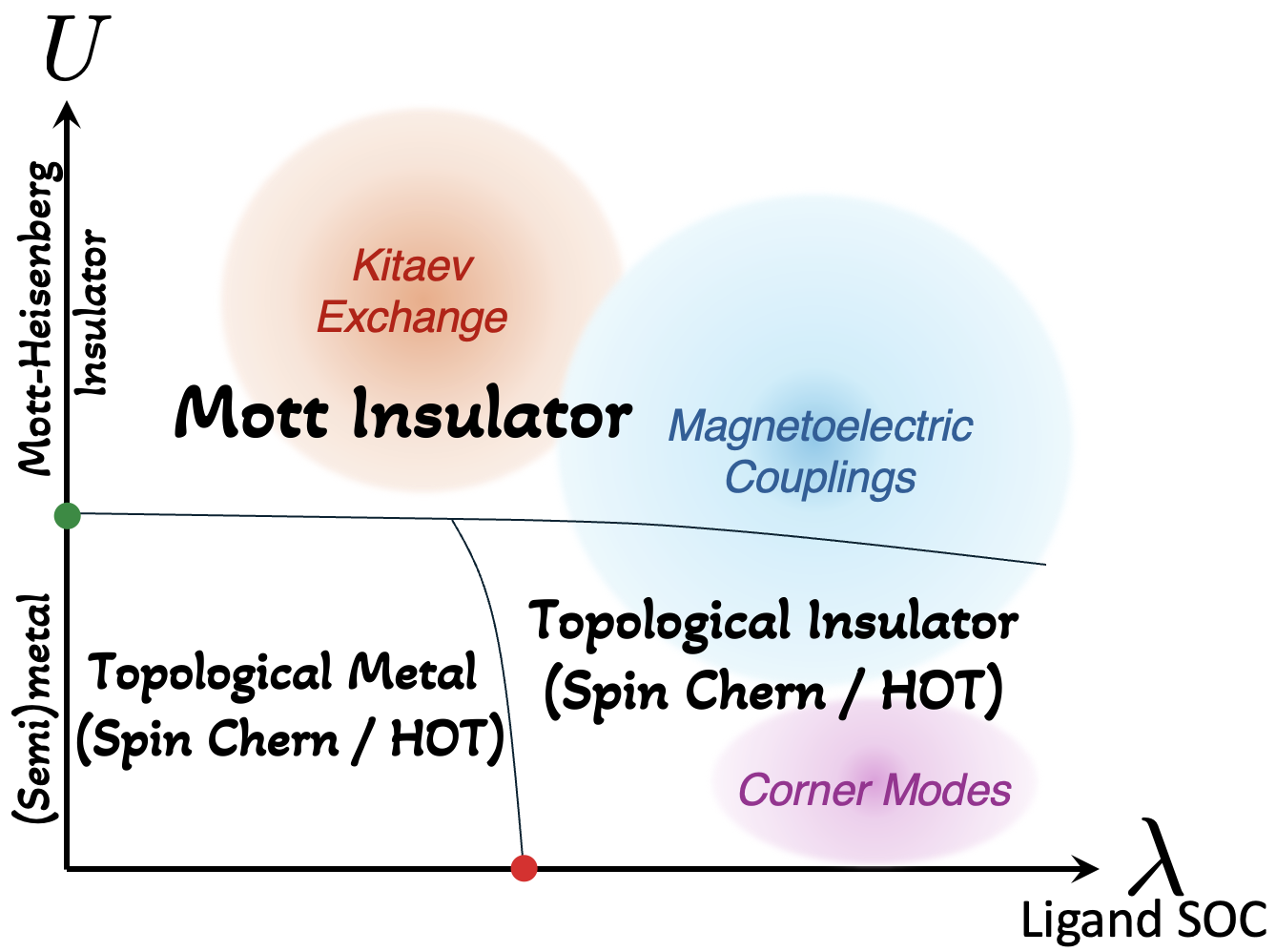}
    \caption{Schematic phase diagram of an effective two-orbital $e_g$ model as function of ligand SOC $\lambda$ and Coulomb correlations $U$ at the metal site. For
    $\lambda=0$, increasing $U$ leads to a transition from a (semi)metal to Mott-Heisenberg insulator. Increasing $\lambda$ 
    for weakly interacting case, leads to topological metal and topological insulator phases featuring spin-Chern bands and higher order topology with 
    corner modes. The large $U$ limit for strong $\lambda$ reveals an intertwining of band topology,
    Kitaev interactions, and 
    magneto-electric effects.}
    \label{fig:schematicphasediag}
\end{figure}

\begin{figure*}[t]
\centering
\includegraphics[width=0.9\textwidth]{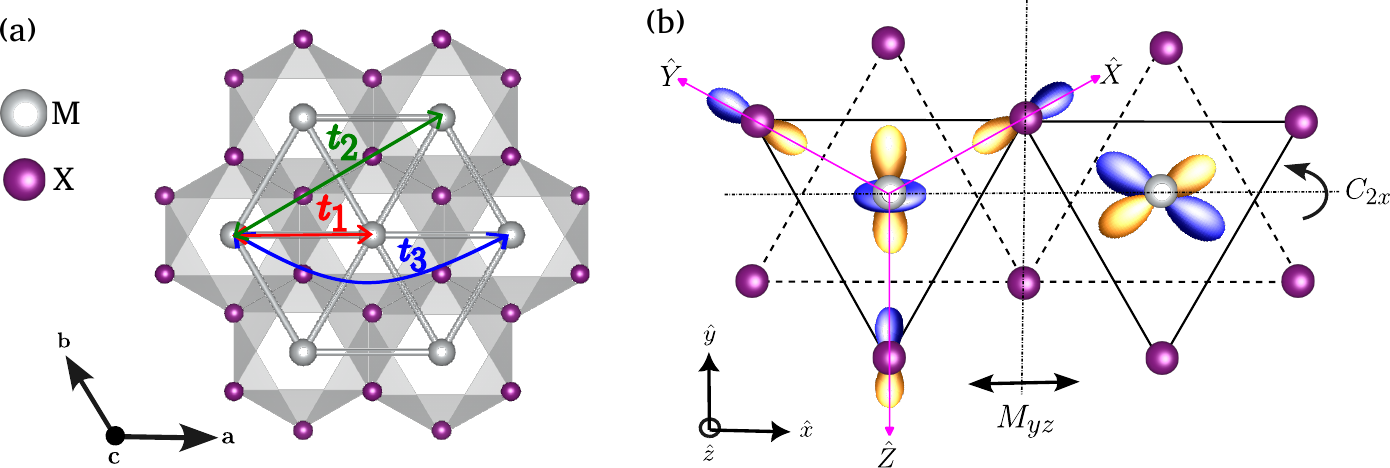}
\caption{(a) Crystal structure of single layer $1T$-Ni$X_2$, showing the edge-shared network of Ni$X_6$ octahedra in crystallographic $\mathbf{a}$-$\mathbf{b}$ plane,
% formed by Ni atom surrounded by ligand X atoms, 
viewed along the $\mathbf{c}$ axis. The effective Ni-Ni hopping terms for first
neighbor $t_1$, second neighbor $t_2$, and third neighbor $t_3$ pairs are marked. (b) Relevant $(d_{3Z^2-r^2},d_{X^2-Y^2})$ orbitals
on Ni sites and ($p_X, p_Y,p_Z$) orbitals on ligand sites, defined in local coordinate systems $\hat{X}$, $\hat{Y}$, and $\hat{Z}$.  In units of the Ni-Ni bond length, the nearest neighbor
Ni-Ni bond vectors are $\vec{\delta}_1 =  \hat{x}, \vec{\delta}_2= -\frac{1}{2}\hat{x} + \frac{\sqrt{3}}{2}\hat{y}$ and $\vec{\delta}_3 = \frac{1}{2}\hat{x} + \frac{\sqrt{3}}{2}\hat{y}$.
where $\hat{x}$, $\hat{y}$, and $\hat{z}$ denote the global Cartesian coordinate axes.
}
\label{fig:drawing_of_NiX2}
\end{figure*}

In this work, motivated by the structure of Ni$X_2$ compounds, which have a triangular net of Ni$^{2+}$ ions enclosed in 
halogen $X^{-}$ octahedra as a case study, we 
build an effective two-orbital $e_g$ model that incorporates the effect of strong SOC at the ligand site through 
spin-dependent hoppings. These $e_g$ orbitals might be viewed as `molecular orbitals' formed on the Ni$X_6$ octahedra.
Although this family of compounds, especially NiI$_2$ has been investigated both theoretically and experimentally \cite{brik2007comparative, miao2025spin, pan2025long, yu2025microscopic, lebedev2024photocurrent}, they have been primarily explored using density functional theory (DFT) calculations~\cite{amoroso2020spontaneous, Fumega_2022,riedl2022microscopic}. 
Furthermore, while downfolding approach has been previously used to construct a five-orbital model, projecting to Ni $d$-orbitals, theoretical work on this complex
model has been limited to a numerical exact diagonalization study of the magnetic exchange interactions in the Mott limit \cite{riedl2022microscopic}.
In comparison, in our work, we discover that the simpler two-orbital effective $e_g$ model already
offers a rich playground for different emergent phases
which can be further probed in the material platform under external perturbations like pressure, strain, and doping.
%is more amenable to obtaining analytic and numerical results. 
 Our primary results, shown schematically in Fig.\ref{fig:schematicphasediag}, are the following.
(i) %As shown schematically in Fig.\ref{fig:schematicphasediag}, 
The non-interacting limit of our model hosts topological bands with nontrivial spin Chern number or higher order topological (HOT) indices. These topological
phases can be metallic or insulating depending on the relative strength of spin-independent hopping terms, and they are
characterized by counterpropagating spin polarized edge modes for the spin Chern bands and 
corner localized modes for the HOT insulator.
(ii) Turning on strong correlations
at the Ni site pushes the system into a Mott insulator. We derive 
analytical results for the resulting bond-dependent Kitaev interactions,
in addition to a Heisenberg exchange, which can stabilize non-trivial spin textures. 
The strong coupling limit of a related model was previously proposed to realize spin-$1$ Kitaev 
magnetism \cite{keePRL}.
(iii) Finally, the non-trivial nature of orbital-dependent spin currents within the two-orbital set-up opens up distinct 
magnetoelectric effects beyond the single-orbital $j_{\rm eff}\!=\! 1/2$ model. We discuss this physics using a bond-polarization 
framework which naturally extends 
previous work on the single-orbital case and also
%our results
goes beyond the generalized Katsura-Nagaosa-Balatsky (g-KNB)
theory \cite{katsura2005spin, xiang2011general}. We uncover hitherto unexplored terms which can play a role when
spins in the two orbitals are not perfectly locked, a situation that can be relevant near the Mott insulator phase boundary. 
%For the specific case of {Ni$X_2$}, we present numerical results based on atomic-limit wavefunctions in the spin-$1$ Mott insulator.
Our study opens up future explorations of the interplay of itinerancy, topology, and strong correlations in $e_g$ orbitals, 
which can be relevant to doping, strain, and
high-pressure studies of Ni$X_2$ and related compounds.

\section{Non-interacting limit}

We begin by constructing the effective $e_g$ model for triangular lattice 1T-$MX_2$ compounds using a combination of symmetry arguments bolstered by 
density functional theory (DFT) insights.
The 2D triangular lattice 1T-$MX_2$ compounds feature edge-sharing {$MX_6$} octahedra, with transition metal ions $M$ experiencing an octahedral crystal field 
from the surrounding $X$ ligands as shown in Fig.\ref{fig:drawing_of_NiX2}. The octahedral crystal field
splits the $d$ orbitals into a lower-energy $t_{2g}$ $(d_{XY}, d_{YZ}, d_{ZX})$ triplet and higher-energy $e_g$ $(d_{X^2-Y^2}, d_{3Z^2-R^2})$ doublet, with
orbitals being defined in the local coordinate systems $\hat{X}$, $\hat{Y}$, and $\hat{Z}$, having an energy splitting of $\approx 1$\,eV.
The space group symmetry of these 2D materials, e.g. $P\overline{3}m1$ for Ni$X_2$,
generically allows for a trigonal distortion of the octahedra. The resulting $D_{3d}$ point group symmetry at the $M$ site 
further splits the $t_{2g}$ manifold into a lower energy $a_{1g}$ singlet and higher energy $e_{g}^\pi$ doublet, with an energy difference 
which is much smaller than the $t_{2g}$-$e_g$ splitting, as well as weak mixing between $t_{2g}$ and $e_g$ states via ligand-metal hybridization. 
Given these typically weak trigonal distortions, 
we will continue to refer to these crystal field split states as $t_{2g}$ and $e_g$ orbitals. 

Hybridization between the metal $e_g$ and ligand $p$ orbitals of the same symmetry leads to the 
formation of $e_g$-like bonding ($e_{g\sigma}$) and antibonding ($e_{g\sigma}^*$) states \cite{haverkort2012multiplet}.
Similarly, $t_{2g}$-$p$ hybridization produces $t_{2g}$-like $\pi$-bonding and $\pi^*$-antibonding states. In Ni$X_2$, with 
nominal $d^8$ configuration of Ni$^{2+}$, 
the antibonding $e_{g\sigma}^*$ level lies close to the Fermi level while $t_{2g\pi}, t_{2g\pi}^*$ and $e_{g\sigma}$ levels lie at significantly lower energy and are fully occupied. This defines the effective $e_g$ degree of freedom in the 
low-energy description of the Ni$X_2$ family. 
In section I of Supplemental Material (SM) \cite{suppmat}, we provide DFT results in generalized gradient approximation (GGA)~\cite{PhysRevLett.77.3865} obtained using a plane-wave basis and projector augmented-wave (PAW) potentials \cite{PhysRevLett.77.3865}, as implemented in the Vienna \textit{Ab initio} Simulation Package (VASP)~\cite{PhysRevB.50.17953,TACKETT2001348,10.1063/1.1926272} in further support of this picture. 

Although the orbital angular momentum for Ni$^{2+}$ should be quenched in the $e_g$ orbital, resulting in negligible SOC effects, 
increasing halogen atomic number and the strong Ni-$e_g$ to ligand-$p$ hybridization 
leads to increasing ligand-mediated SOC, making it an interplaying field of multi-orbital physics and bond-dependent SOC.
In the following, we work out the non-interacting $e_g$ model that incorporates the ligand SOC effect, and uncover its non-trivial topological properties.

\begin{figure*}[t]
    \centering
    \includegraphics[width=0.9\linewidth]{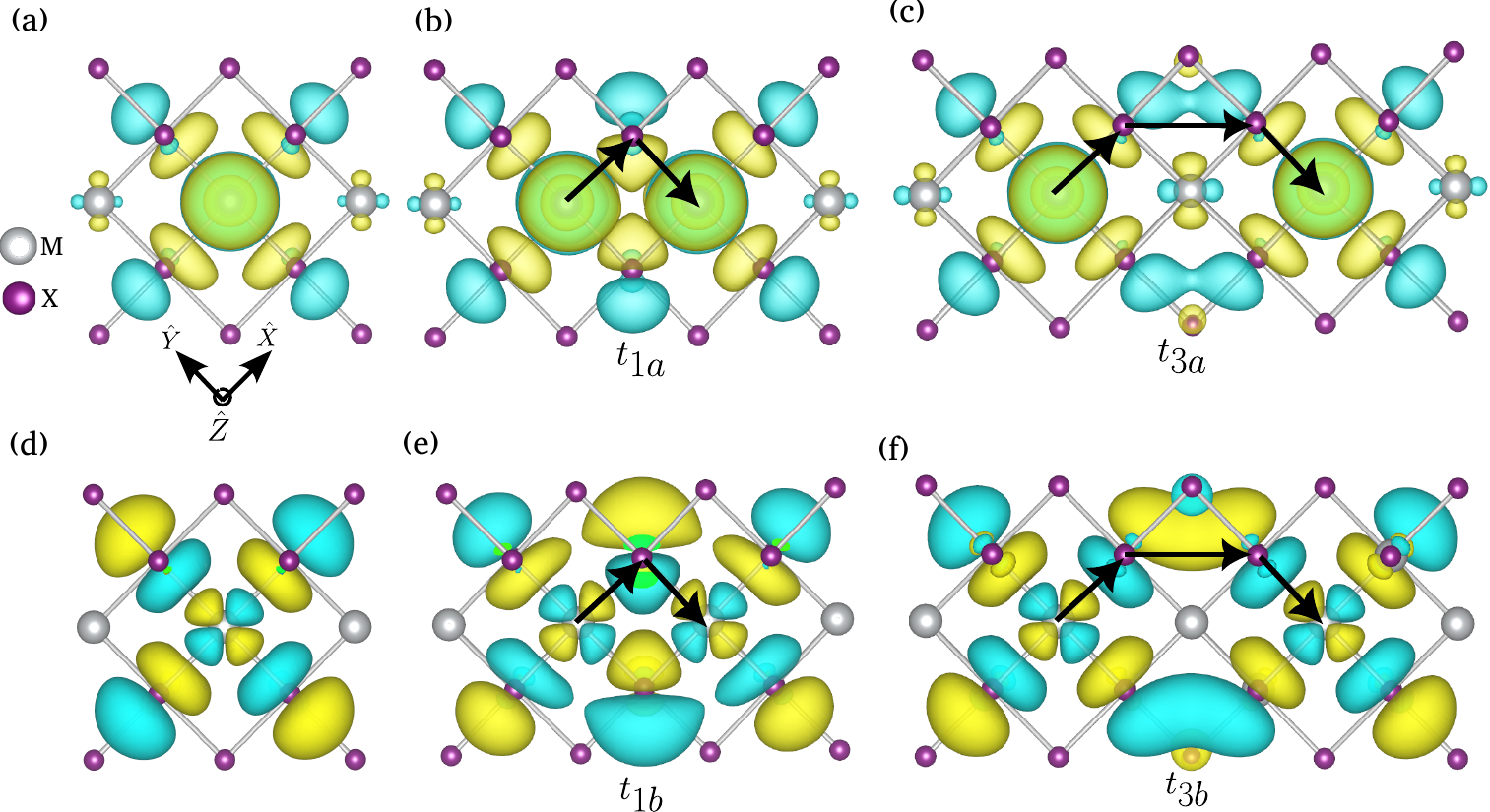}
    \caption{DFT-derived $e_g$ Wannier functions, and their overlaps defining ligand-mediated spin-independent hopping pathways for $t_1$ and $t_a$, for the representative case of NiI$_2$. 
    The panels (a)-(c) show the Ni centered $d_{3Z^2-r^2}$ Wannier function, their overlap at two nearest neighbor Ni sites, defining $t_1$ and their overlap at two third neighbor Ni sites, defining $t_3$, respectively. Panels (d)-(f) show the same but for $d_{X^2-Y^2}$ Wannier functions. Plotted are the isovalue surface of the functions with different signs of the lobes colored differently. The ligand mediated hopping paths are shown with arrows.}
    \label{fig:wannier_orbitals}
\end{figure*}

\subsection{Two-orbital tight-binding model including ligand SOC: Symmetry constraints}

Focusing on the nearest-neighbor (NN) $\vec{\delta}_1$-bond, i.e., the bond along the $x$-axis (cf Fig.\ref{fig:drawing_of_NiX2}),
the hopping matrix is constrained by the mirror symmetry $M_{yz}$ at the bond center, a two-fold rotation $C_{2x}$ along the bond, and time-reversal symmetry $\mathcal{T}$. The most general symmetry-allowed tight-binding Hamiltonian on this bond is
\begin{gather}\label{NN electron hopping matrix along x}
\begin{aligned}
&\hat{T}_{(i,i+\vec{\delta}_1)} = \Psi_i^\dag T^\pdg_{i,i+\delta_1} \Psi_{i+\delta_1}^\pdg + \Psi_{i+\delta_1}^\dag T_{i,i+\delta_1} \Psi_i, \\
 % \Psi_i^\dag \begin{bmatrix} -t_{1a} &-i\lambda_{1z} &0 &\lambda_{1y} \\ i\lambda_{1z} &-t_{1b} &-\lambda_{1y} &0 \\ 0 &-\lambda_{1y} &-t_{1a} &i\lambda_{1z} \\ \lambda_{1y} &0 &-i\lambda_{1z} &-t_{1b} \end{bmatrix} \Psi_{i+\vec{\delta}_1}, \\
\text{where} \quad &T^\pdg_{i,i+\delta_1} = \begin{bmatrix} \!- t_{1a} &0 \\ 0 &\! - t_{1b}\end{bmatrix} \sigma_0 \!+\! 
\lambda_{1z} \tau_y \sigma_z \!-\! \lambda_{1y} \tau_y \sigma_y.
\end{aligned}
\end{gather}
where $\Psi_i^\dag = [a_{i\ua}^\dag, b_{i\ua}^\dag, a_{i\da}^\dag, b_{i\da}^\dag]$ denotes the $e_g$ creation operators on site $i$, $d_{3Z^2-r^2}$ and $d_{X^2-Y^2}$ being labelled
as $a$ and $b$ orbitals. The spin and orbital Pauli matrices are denoted by $\sigma$ and $\tau$ respectively (with $\sigma^0,\tau^0$ denoting corresponding identity matrices). 
For convenience, we take the spin operators to be aligned with the 
global $(x,y,z)$ axes shown in Fig.~\ref{fig:drawing_of_NiX2}. 
Here $\lambda_{1z}$ and $\lambda_{1y}$ are real parameters corresponding to effective spin-preserving and spin-mixing SOC-induced hoppings 
between $e_g$ orbitals originating from the strong SOC of the intervening ligand $p$-orbitals.
% and the $p\text{-}d$ hybridization, as shown in SM \cite{suppmat}.
The hopping matrices on other bonds can be obtained by applying the 3-fold rotation $C_{3z}$.

 The same symmetry constraints discussed above then fixes the form of the 2NN and 3NN hopping matrix along $(\delta_1+\delta_2)$ and $2\delta_1$ (cf Fig.\ref{fig:drawing_of_NiX2}), respectively to be
\begin{align}
\label{3NN electron hopping along x}
T_{i,i+(\vec{\delta}_1+\vec{\delta}_2)}  &=  \begin{bmatrix} \!- t_{2a} &0 \\ 0 &\! - t_{2b}\end{bmatrix} \sigma_0  + 
\lambda_{2z} \tau_y \sigma_z - \lambda_{2y} \tau_y \sigma_y, \\
T_{i,i+2\vec{\delta}_1} \! &= \begin{bmatrix} \!- t_{3a} &0 \\ 0 &\! - t_{3b}\end{bmatrix} \sigma_0 + 
\lambda_{3z} \tau_y \sigma_z - \lambda_{3y} \tau_y \sigma_y.
\end{align}
The ligand-$p$-mediated effective Ni-Ni NN, 2NN and 3NN hoppings arises from 
nearest-neighbor Ni-$X$ hoppings, and nearest-neighbor $X$-$X$ hoppings. As shown in section II of SM \cite{suppmat}, for the 2NN hoppings 
along the $(\delta_2+\delta_3)$-bond, the intralayer $X$-$X$ paths add destructively, while only one interlayer $X$-$X$ path contributes to the spin-independent hoppings, yielding
suppression of the spin-independent 2NN hoppings ($t_{2a}$, $t_{2b}$) compared to the 1NN and 3NN terms. Also, this makes spin-dependent 2NN ($\lambda_{2z}$, $\lambda_{2y}$) and 
3NN hoppings ($\lambda_{3z}$, $\lambda_{3y}$) significantly smaller compared to $\lambda_{1y}$ and $\lambda_{1z}$. 
The minimal two-orbital tight-binding model thus consists of spin-independent NN and 3NN hoppings, $t_{1a}, t_{1b}$ and $t_{3a}, t_{3b}$, and spin-preserving and spin-mixing SOC-induced NN hoppings, $\lambda_{1z}$ and $\lambda_{1y}$. The full  momentum
space Hamiltonian is explicitly given in section II.A of SM \cite{suppmat}, along with representative plots of the dispersion
without and with SOC terms.

% \textcolor{red}{Discuss issue of why naive atomic-level PT/DF fails for NiI2.}
% of the Ni $e_g$-$X$ $p$ model including Ni-$X$-Ni and Ni-$X$-$X$-Ni paths

\subsection{Hierarchy of hopping energy scales: Insights from DFT}

To gain insight into the relative strengths of $t_{1a}$, $t_{1b}$, $t_{3a}$, and $t_{3b}$, 
we construct the maximally localized Wannier functions, starting from a self-consistent all orbital 
DFT calculation, and integrating out all degrees of freedom except Ni $e_g$. This yields effective Ni-centered $e_g$ Wannier functions, as shown in Fig.\ref{fig:wannier_orbitals} for the representative case of NiI$_2$. While the central part of these effective Wannier functions are shaped according to $d_{3Z^2-r^2}$ or $d_{X^2-Y^2}$ symmetry, the tails sitting at the neighboring $X$ sites are shaped according to
$X$-$p_x$ or $X$-$p_y$ symmetries, indicating the formation of strong pd$\sigma$ antibonds.

For nearest-neighbor Ni pairs, the overlap of the tails of the Wannier functions at the common edge-shared $X$ positions is almost
orthogonal, thereby significantly weakening the effective NN hopping. On the other hand, for the Wannier functions placed at 3NN Ni sites, the tails at Ni sites of the two 
functions point towards each other (marked by arrows in Fig. 3), forming a strong connecting $X$-$X$ pathway of the two functions. This rationalizes $t_3$ being larger
than $t_1$. Within the 2D planar geometry,  bonding of I-$p_x$/I-$p_y$ orbitals is stronger with the $d_{X^2-Y^2}$ orbital compared to the $d_{3Z^2-r^2}$ orbital, making
the ligand-mediated hopping pathways much more effective for $d_{X^2-Y^2}$ (``{\it b}'' orbital) compared to $d_{3Z^2-r^2}$ (``{\it a}'' orbital). This makes $t_b$ $>$ $t_a$ and fixes the hierarchy of hoppings $\vert t_{3b} \vert > \vert t_{3a} \vert > \vert t_{1b} \vert > \vert t_{1a} \vert $.
The ratios $\vert t_{1a} \vert/\vert t_{3b} \vert$, as well as $\vert t_{1b} \vert/\vert t_{3b} \vert$, and
$\vert t_{3a} \vert/\vert t_{3b} \vert$ depend on
the charge-transfer energies $(\varepsilon_{e_g} - \varepsilon_p)$, Ni-$X$ hopping strength $t_{pd\sigma}$, $X$-$X$ hopping strength ($t_{pp\sigma}$, $t_{pp\pi}$) as shown in section II.B of SM \cite{suppmat},
and thus shows material dependency on change of $X$ from I (5p) to Br (4p) to Cl (3p). This trend is reflected in the parameters listed in Table \ref{DFT parameters}.

Additionally, we extracted the SOC coupling terms, $(\lambda_{1z},\lambda_{1y})$, by fitting the tight-binding dispersion to the DFT 
band structure obtained including SOC. 
The extracted SOC strengths are also listed in Table~\ref{DFT parameters}. 
In the limit where the ligand SOC $\lambda_p \ll \Delta_{*}$, where $\Delta_{*}$ is the energy separation between
the antibonding $e_{g\sigma}*$ and ligand-$p$ level, the leading order perturbative derivation leads to $\lambda_{1y} \approx
\sqrt{2} \lambda_{1z}$ \cite{suppmat}; while this relationship roughly
holds for NiCl$_2$ and NiBr$_2$, there is significant deviation for NiI$_2$.
We find that this simple relation between $\lambda_{1y}$ and $\lambda_{1z}$ is generally 
violated once longer-range $X$-$X$ hoppings are included (see section II.B of SM \cite{suppmat}). 
% More importantly, the strong Ni-$X$ covalency and smaller energy difference between Ni-$e_g$ and ligand 
% $p$-orbitals for iodine leads to a breakdown of
% naive perturbative downfolding to atomic-like orbitals in an effective $dp$ model for {NiI$_2$}. \textcolor{red}{TSD: Is this last line really needed ? It sounds like it challenges eg-model
% itself for NiI2.}

\begin{table}[t]
\begin{centering}
\begin{tabular}{|l|l|l|l|l|l|l|}
\hline
 Parameters (meV) &~$t_{1a}~$ &~$t_{1b}$~  &~$t_{3a}$~  &~$t_{3b}$~  &~$\lambda_{1z}$~ &~$\lambda_{1y}$~\\ \hhline{|=|=|=|=|=|=|=|}
~~~$\text{NiCl}_2$   &$-4$     &$-2$  &$12$  &$-58$  &$2$  &$3$  \\ \hline
~~~$\text{NiBr}_2$   &$-2$     &$-19$  &$15$  &$-69$  &$7$ &$10$  \\ \hline
~~~$\text{NiI}_2$    &$-22$     &$-36$  &$25$  &$-95$  &$14$ &$28$ \\ \hline
\end{tabular}
\end{centering}
\caption{Hopping parameters and SOC terms in the effective two-orbital model for Ni$X_{2}$ obtained from the maximally localized the 
Wannier-basis sets in DFT.}
\label{DFT parameters}
\end{table}

\subsection{\label{sec:topo}Topological properties of the two-orbital model}

The presence of spin-preserving and spin-flip hoppings in the effective tight-binding model, arising from the 
SOC of the ligand, opens up the
possibility of non-trivial band topologies. Indeed, as demonstrated in the following subsections, 
the effective two-band model with interplay of $\lambda_{1z}$ and $\lambda_{1y}$ is found to
display both nontrivial first-order topology characterized by non-trivial spin Chern number and and higher-order topology 
with corner modes as shown in Fig.~\ref{Spin-Chern Phase diagram}.
Higher-order topology has been previously reported via different routes in the 
broader class of {1T-M$X_2$} materials based on first-principles
calculations Ref.~\cite{long2024second,arroyo2025two}. In particular, Ref.~\cite{long2024second} identified HOT in monolayer $\mathrm{1T-PtSe_2}$ material class, and showed, using an effective edge theory, that the resulting corner states are protected by crystalline symmetry. Ref.~\cite{arroyo2025two} carried out a related analysis and further interpreted the higher-order topology as arising from an orbital-mediated obstructed atomic limit.  Notably, these mechanism do not rely on SOC. In contrast, our results discussed below
complement these studies by providing an effective $e_g$-orbital perspective and demonstrating the emergence of SOC-driven corner states. HOT in our model arises between the interplay of first-order topology and the spin-mixing $\lambda_{1y}$ term through a mass-kink mechanism \cite{hung2024time}. This is reminiscent of Zeeman-field induced corner states in first-order topological insulators \cite{ren2020engineering,zhuang2022topological,liu2024two}.

% Moreover, the same analysis can be extended to $t_{2g}$ manifold, which is relevant for a broader class of materials within this family.

\subsubsection{Spin-Chern bands  \\ ($\lambda_{1z} \ne 0$, $\lambda_{1y} = 0$)}

\begin{figure}[t]
\centering
\includegraphics[width=0.48\textwidth]{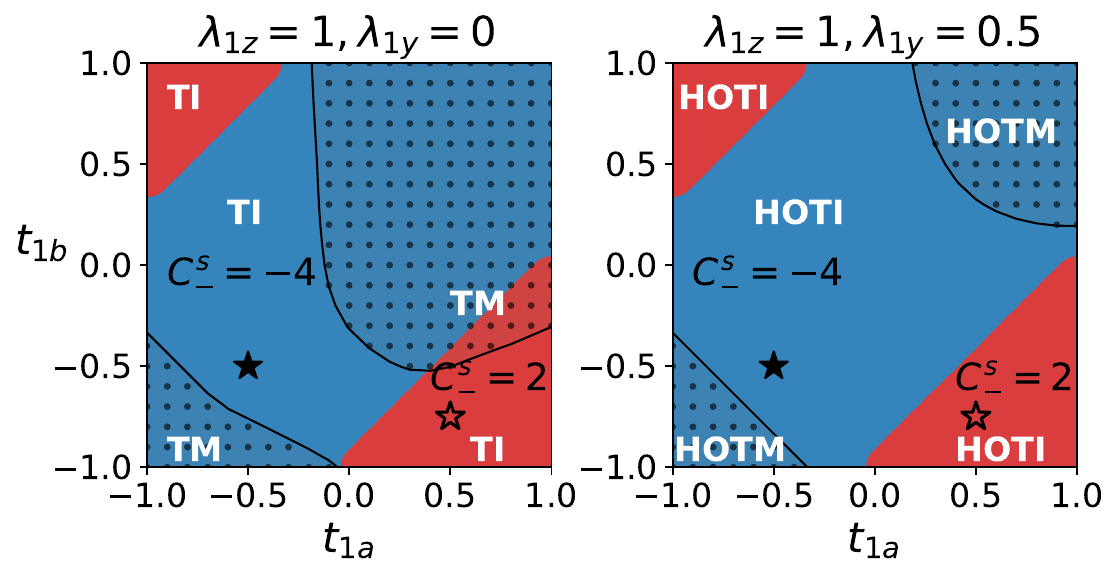}
\caption{Phase diagram showing the various topological phases 
$C^{s}_{-}$ for the lower band as a function of the NN hoppings $t_{1a}, t_{1b}$. The 3NN hopping parameters are fixed to $(t_{3a}, t_{3b}) = (1/3, -1)$, and we take $\lambda_{1z} >0$. The blue and red regions corresponds to $C^{s}_{-} = -4$ and $C^{s}_{-} = 2$, respectively. The dot-shaded region indicates metallic phases, while the unshaded region corresponds to insulating phases, and the solid black line is the metal-insulator boundary. We label phases as topological insulating (TI) and topological metallic (TM), and the higher
order analogues as HOTI and HOTM. The bulk and edge spectra corresponding to the solid stars $(C_-^s = -4)$ in each panel are shown in Fig.~\ref{Example of bulk, edge state with Cs = -4}. The corresponding results for the empty stars $(C_-^s = 2)$ are presented in SM
Section III.A \cite{suppmat}.}
\label{Spin-Chern Phase diagram}
\end{figure}

We first study the case $\lambda_{1y} = 0$, which conserves $S_z$, and vary $\lambda_{1z}$ to access different phases.
At $\lambda_{1z} = 0$, the Bloch spectrum of the tight-binding model exhibits a metallic phase with
band touchings at the $\Gamma$ and $K$ points (see section II.A of SM \cite{suppmat}).
% as shown in Fig.\ref{Spin-Chern Phase diagram}.
For $\lambda_{1z} \neq 0$, the degeneracy at these band touching points is lifted
and the spectrum becomes (locally) gapped at every momentum $\vec{k}$, as plotted in 
Fig.~\ref{Example of bulk, edge state with Cs = -4}(a) for a choice $t_{3a}$ = 1/3 and $t_{3b}$ = -1;
we note, however, that this is the generic scenario. 
In this case, the spectrum is comprised of a pair of isolated two-fold degenerate bands, denoted
as upper ($+$) and lower ($-$) bands. The band degeneracy is protected by the
combined presence of time-reversal and inversion symmetries.
Since $S_z$ is conserved, one can define projectors $P_{\pm}^{\ua, \da}(\vec{k})$ onto the $+$ and $-$ bands 
in the spin sectors $\ua, \da$. Focusing on the lower $-$ band, we compute the Chern 
number $C^{\sigma}_{-}$ for each spin sector $\sigma = \ua, \da$, noting that for the upper band $C^{\sigma}_{+} = -C^{\sigma}_{-}$.
We then define the spin-Chern number as $C^s_{-} = \frac{1}{2}(C^{\ua}_{-} - C^{\da}_{-})$. 
Within this model, depending on choice of values of $t_{1a}$ and $t_{1b}$, $C^s_{-}$ can take nontrivial even integer values. For the choice $t_{3a}/t_{3b} = -1/3$, which is close to the DFT extracted ratio for the {Ni$X_2$} compounds, 
% as shown in Fig.\ref{Spin-Chern Phase diagram} 
the topological phases for $\lambda_{1z} > 0$ exhibit two distinct values of the spin Chern number, 
$C_{-}^{s} = -2$ and $C_{-}^{s} = 4$ as shown in Fig.~\ref{Spin-Chern Phase diagram}(a). 
 The sign of $C^{s}_{-}$ is controlled by the sign of $\lambda_{1z}$: reversing $\lambda_{1z}$ flips the sign of $C^{s}_{-}$, leaving the phase boundaries unchanged. 
An odd spin-Chern number is 
forbidden since even-parity $e_g$ orbitals restricts us to trivial $\mathbb{Z}_2$ index \cite{fu2007topological}.

\begin{figure}[t]
% \centering
% \subfloat[$C_s = 2$]{\includegraphics[width=0.47\textwidth]{Figures/Topology_Figures/Cs_2_edge_gapped_example.pdf}}
% \label{Example of bulk, edge state with Cs = 2}
\includegraphics[width=0.49\textwidth]{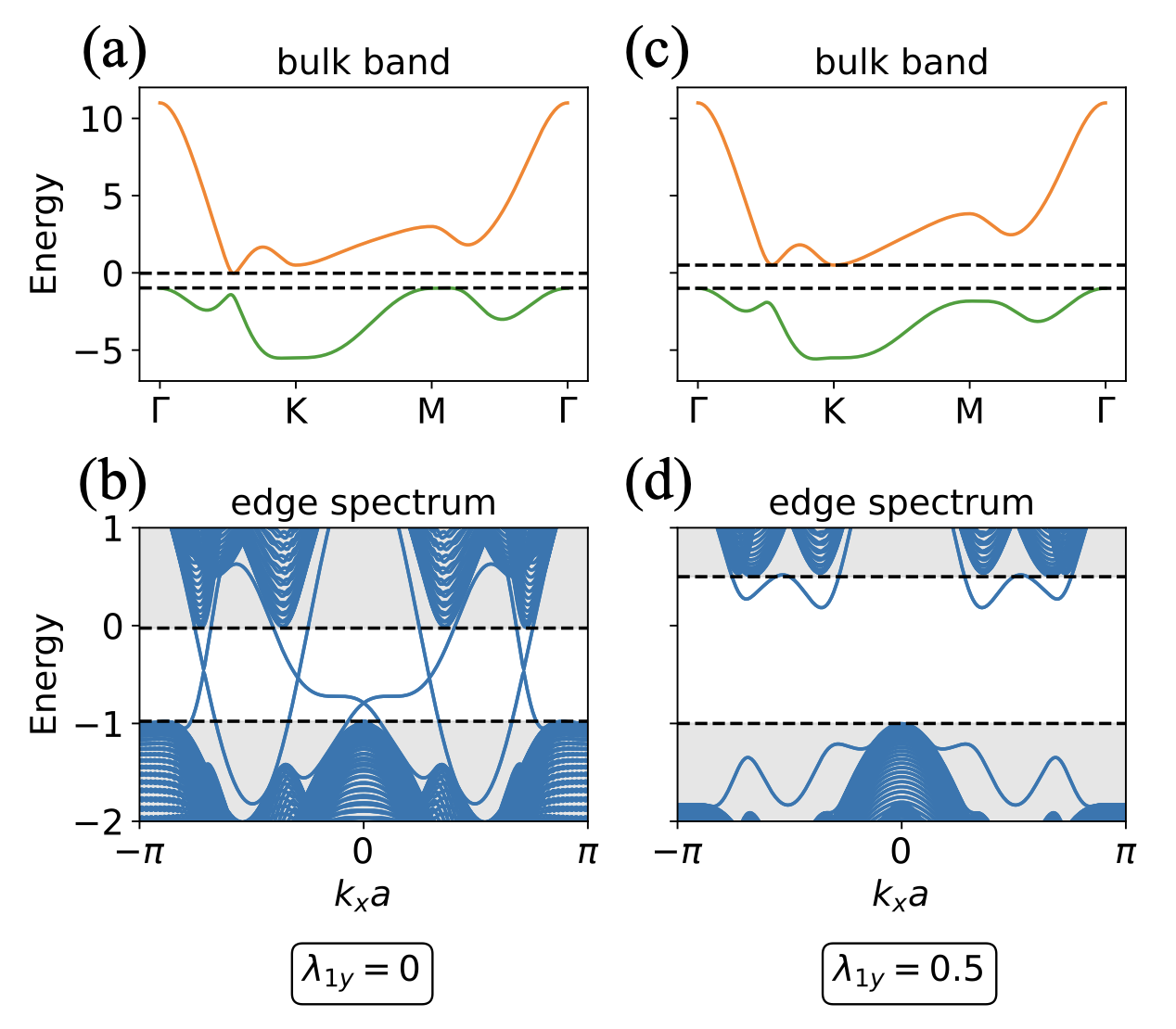}
\caption{Gapped bulk band dispersion, showing $+$ and $-$ bands colored as orange and green respectively, and edge 
spectrum for $C^s_{-} = -4$ for spin-mixing SOC $\lambda_{1y}=0$ (left) and
$\lambda_{1y}=0.5$ (right). A nonzero $\lambda_{1y}$ enhances the bulk band gap of the topological insulator,
and gaps out the edge states. Analogous results for the regime of the phase diagram with $C^s_{-} = 2$ are
presented in SM Section III.A \cite{suppmat}}
\label{Example of bulk, edge state with Cs = -4}
\end{figure}

A nonzero spin-Chern number $C_{-}^{\sigma}$ indicates a nontrivial bulk topology, and hence the presence of the associated gapless edge states. As shown in Fig.\ref{Example of bulk, edge state with Cs = -4}, for $(t_{1a}, t_{1b}) = (-0.5,-0.5)$ 
yielding $C^{s}_{-} = -4$, the bulk spectrum is completely gapped and we find four pairs of spin-polarized gapless edge modes 
consistent with a quantum spin Hall (QSH) insulator. In the SM \cite{suppmat}, we show similar results for the case of
$C^{s}_{-} = +2$.

The phase diagram in $t_{1a}$-$t_{1b}$ plane also exhibits metal-insulator transitions that are of non-topological 
nature. In
the topological metal (TM) phase (indicated by a dotted region), the chemical potential necessarily 
crosses both the dispersive spin-Chern bands leading to metallic Fermi surfaces in the bulk.
In the topological insulator (TI) phase, the Fermi level lies between well separated spin-Chern bands leading to a 
quantum spin Hall (QSH) insulator at half-filling.

% Fig.~\ref{Example of bulk, edge state with Cs = 2, -4}(b) shows the edge spectrum for the $C_{-}^{s} = 4$. We find fully
% gapped bulk bands and gapless helical edge states .

% As demonstrated in Fig.\ref{Spin-Chern Phase diagram}, the value of $\lambda_{1z}$ shifts the MIT lime: increasing $|\lambda_{1z}|$ expands the insulating regime by contracting the metallic regime. 
% By contrast, the sign of spin Chern number $C^{s}_{-}$ is controlled only by the sign of $\lambda_{1z}$: reversing $\lambda_{1z}$ flips the sign of $C^{s}_{-}$, leaving the phase boundaries unchanged. 

\subsubsection{Higher Order Topology \& Corner Modes \\ ($\lambda_{1z} \ne 0$, $\lambda_{1y} \neq 0$)} \label{Section: Corner states}

We next consider turning on $\lambda_{1y} \neq 0$. In this case, as shown in 
Fig.~\ref{Example of bulk, edge state with Cs = -4}(c) and (d), the bulk gap remains nonzero and even gets slightly enhanced compared to $\lambda_{1y} = 0$,
while the gapless edge modes get gapped out. The resulting insulator
exhibits corner states within the edge gap\cite{hung2024time}, stabilizing
a phase with nontrivial higher order topology\cite{ren2020engineering,zhuang2022topological,liu2024two}.

To identify the corner modes, we employ a finite hexagonal flake that preserve both $C_3$-symmetry and inversion symmetry $\mathcal{I}$, though the existence of corner mode does not depend on the specific shape. In this geometry, the HOT is directly connected to a filling anomaly and can be characterized by symmetry indicators \cite{benalcazar2019quantization, schindler2019fractional}.
% Upon inclusion of spin-mixing $\lambda_{1y}$ and gapping out of the edge states, 
As seen from Fig.~\ref{Example of corner state with Cs = 2, -4}(a), the spectrum exhibits $12$ nearly degenerate in-gap 
levels; from the wavefunctions in
Fig.~\ref{Example of corner state with Cs = 2, -4}(b), we find that we can identify these levels with localized
corner modes. At half-filling, $8$ levels are filled, which leads to a fractional corner charge
$\mathcal{Q}_c = 4e/3 \text{ mod } 2e$ per corner of the hexagonal flake.

% \begin{widetext}
\begin{figure}[h]
% \centering
% % \subfloat[$C_s = 2$]
% \includegraphics[width=0.47\textwidth]{Figures/Topology_Figures/Cs_2_corner_states_example.pdf}\label{Example of corner state with Cs = 2} 
% % \hfill
% % \subfloat[$C_s = -4$]
\includegraphics[width=0.47\textwidth]{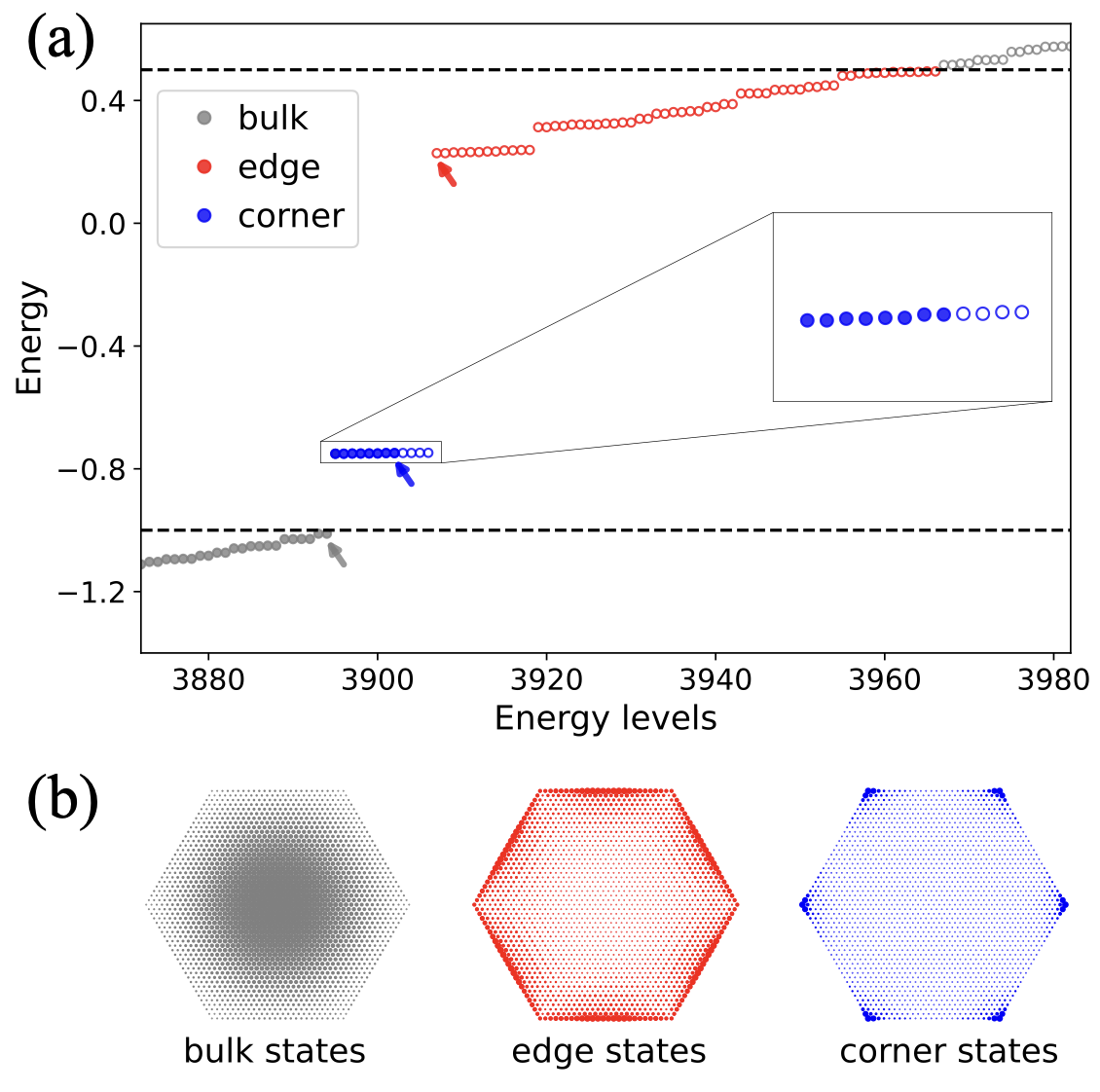}\label{Example of corner state with Cs = -4}
% \vspace{0.8em}
% \subfloat[$|\Psi(i)|^2$]
% \includegraphics[width=0.7\textwidth]{Figures/Topology_Figures/Cs_states_distribution.png}\label{states distribution} 
\caption{Top panel: Energy spectrum of a finite $C_3+\mathcal{I}$-symmetric hexagonal flake with $1951$ sites. Filled and unfilled states are 
indicated by closed and open circles, respectively, and the color distinguishes the bulk states (gray), edge modes (red) and corner modes
(blue). The 
inset zooms into the states near the Fermi level to highlight the fractional filling of
12 degenerate corner modes. 
% Panels (a) and (b) 
% correspond to $C_s = 2$ and $C_s = -4$, respectively. 
Bottom panel: Real-space charge distribution $|\Psi(i)|^2$ of 
the states indicated by arrows in top panel to highlight their
bulk, edge, and corner nature.}
% The size of the circles indicates the local charge density.}
\label{Example of corner state with Cs = 2, -4}
\end{figure}

The HOT in this model is protected by rotational and inversion symmetries, $C_3+\mathcal{I}$, and 
we can express the fractional corner charge
in terms of symmetry indicators $\chi_{\mathcal{I}}^{(3)} = \{\big[ M_2^{\mathcal{I}}\big], \big[ K_2^{(3)}\big]  \}$ as
\begin{gather}
\mathcal{Q}_c = -\frac{1}{4}\big[ M_2^{\mathcal{I}}\big] - \frac{1}{3}\big[ K_2^{(3)} \big] \quad \text{mod } 2.
\end{gather}
We find $\chi_{\mathcal{I}}^{(3)} = \{0, -2\}, \mathcal{Q}_c = 4e/3$ for $\lambda_{1z} >0$, while $\chi_{\mathcal{I}}^{(3)} = \{0, 2\}, \mathcal{Q}_c = 2e/3$ for $\lambda_{1z} <0$. Thus, the fractional corner charge depends only on the sign of $\lambda_{1z}$ and is independent of the precise values of $(t_{1a}, t_{1b}, t_{3a}, t_{3b})$ 
which control the value of the spin-Chern number $C^{s}_{-}$, and the spin-mixing $\lambda_{1y}$. 
This is in agreement with our numerical results.

In the above example, we started from a topological (QSH) insulator at $\lambda_{1y} = 0$ and then turned on $\lambda_{1y} \neq 0$.
% considered a parameter regime in which the bulk spectrum is already gapped at $\lambda_{1y} = 0$, with gapless edge Dirac cone appearing within the bulk gap. 
Turning on $\lambda_{1y}$ gaps out the gapless edge modes, leading to corner states within the resulting edge gap \cite{hung2024time}. 
By contrast, we can also start with a topological metal at $\lambda_{1y}=0$. In such cases, the gapless edge states due to the 
spin Chern bands overlap (in energy) with the bulk band state, so the resulting corner states are embedded in the bulk continuum. In this case, even if increasing $\lambda_{1y}$ results in a HOT insulator, the corner states are not guaranteed to lie within the bulk gap unless additional protecting symmetries are present. This motivates the question of whether a well-defined topological indicator can still diagnose higher-order topology in this regime.

To address this question, we follow Ref.~\cite{peterson2020fractional} and employ the fractional corner anomaly (FCA), a band-resolved topological indicator defined without explicit reference to electronic filling or charge neutrality. For a given band, the FCA is defined as
\begin{gather}
\phi = \rho_{\text{corner}} - (\sigma_1 + \sigma_2) \quad \text{ mod 2e},
\end{gather}
where $\rho_{\text{corner}}$ denotes the charge density localized at a given corner, and $\sigma_1 + \sigma_2$ represents the corresponding edge-induced contribution at the same corner. Both quantities are evaluated by filling the entire band.
To illustrate its relation to HOT and its connection with the fractional corner charge $\mathcal{Q}_c$, we consider the parameter set $(t_{1a}, t_{1b}, t_{3a}, t_{3b}, \lambda_{1z}, \lambda_{1y}) = (0, 0.3, 1/3, -1, 1, 0.5)$ for the calculations. 
Diagonalizing on a finite hexagonal flake with $N_{\text{site}} = 1951$, we obtain two well-separated energy bands with a 
gap: bulk valence and conduction
bands with corner states submerged within the valence band spectrum (see SM Section III.B for figure showing spectrum).
% \textcolor{blue}{(see Fig.~S5 in SM \cite{suppmat})}.
For the lower band $(-)$, we obtain $\rho_{\text{corner}} = 0.581\text{e}$ and $\sigma_1 + \sigma_2 = 0.011\text{e}$, yielding an FCA $\phi_{\text{band $(-)$}} = 0.570\text{e} \approx 2\text{e}/3$.
Comparing with the fractional corner charge obtained from symmetry indicators, we find 
$\phi_{\text{band $(-)$}} \approx 2\text{e}/3 =-4\text{e}/3 \text{ mod 2e} =-\mathcal{Q}_c$. 
Importantly, $\phi_{\text{band $(-)$}}$ takes a robust fractional value  even though the corner states are not spectrally isolated, demonstrating that the higher-order topology is robustly encoded in the bulk band. Since the FCA corresponds to the local density of states integrated over the entire band, it may, in principle, be accessible via scanning tunneling spectroscopy.

We, however, note that relatively large spin-dependent hopping amplitudes $(|\lambda_{1y}/t_{3b}|$ and $|\lambda_{1z}/t_{3b}|)$, are required to gap out the 
nonmagnetic band structure. Moreover, many compounds in this class exhibit Mott insulating phases with magnetic order, which may 
compete with or suppress the SOC-driven higher-order topology. 
Therefore, higher-order topological bands are more likely to be realized in the $\mathrm{NiX_2}$ family under doping,
strain, or pressure, where Coulomb interactions are effectively screened and magnetic order is suppressed.

\section{Interacting Limit}
Having explored the rich band topologies of the $e_g$ two-orbital model, we next turn to the impact of interactions within this model. Deep in the Mott insulator limit at half-filling, Hund's coupling leads to a spin-1
moment and we calculate the effective exchange interactions between these $S=1$ spins. We also study magnetoelectric effects in
this two-orbital model, generalizing the relation between
symmetry dictated polarization and spin currents to the multiorbital framework. 

% We next consider the large-U regime of the our spin-orbit coupled tight-binding model. In particular, we explore the magnetism of the S=1 Mott insulating phase, formed by half-filled $e_g$ orbitals
% under strong Hund's coupling, and the spin-induced polarization within the multi-orbital framework. 

\subsection{Mott Magnetism} 

% In particular, sizable Kitaev interactions in these materials have been proposed based on perturbative analyses involving both metal-$d$ and ligand-$p$ orbitals \cite{stavropoulos2019microscopic}, and have been further studied through extensive first-principle calculations \cite{amoroso2020spontaneous, riedl2022microscopic, li2023realistic} have been studied. However, these results have not yet been discussed within the framework of an effective two-orbital model, which we address in the following.

To derive the $S=1$ spin model for half-filled $e_g$ orbital, we start from the local Hubbard-Kanamori Hamiltonian (see section IV of SM \cite{suppmat}), parametrized by the intraorbital repulsion 
$U$, interorbital repulsion $U^{'}$, Hund's coupling $J_H$ and  pair-hopping $J^{'}$. In the strong-coupling limit $U,U^{'},J_H, J^{'} \gg t_{1\alpha}, t_{3\alpha}, \lambda_{1z}, \lambda_{1y}$, 
the low-energy subspace on each site is the $S = 1$ triplet. To connect with the extended Heisenberg-Kitaev $\mathrm{J-K-\Gamma-\Gamma'}$ parametrization \cite{winter2017models, riedl2022microscopic}, we write the bond-dependent interaction as
\begin{gather}\label{JKGG' model}
\begin{aligned}
H_{ij} = &J_{ij} \vec{S}_{i} \cdot \vec{S}_{j} + K_{ij} S_{i}^{\gamma} S_{j}^{\gamma} + \Gamma_{ij} \big( S_{i}^{\alpha} S_{j}^{\beta} + S_{i}^{\beta} S_{j}^{\alpha} \big) \\
&\, + \Gamma_{ij}'\big( S_{i}^{\gamma}S_{j}^{\alpha} + S_{i}^{\gamma}S_{j}^{\beta} + S_{i}^{\alpha}S_{j}^{\gamma} + S_{i}^{\beta}S_{j}^{\gamma}\big) , 
\end{aligned}
\end{gather}
where $\{\alpha, \beta, \gamma \} = \{X,Y,Z \}, \{ Y,Z,X\}, \{Z,X,Y\}$ for bond along $\vec{\delta}_1 =  a\hat{x}, \vec{\delta}_2= -\frac{1}{2}a\hat{x} + \frac{\sqrt{3}}{2}a\hat{y})$ and $\vec{\delta}_3 = \frac{1}{2}a\hat{x} + \frac{\sqrt{3}}{2}a\hat{y}$, respectively. The exchange parameters from perturbation theory\cite{suppmat} are given by,
\begin{gather}\label{JKGG' parameters from perturbation}
\begin{aligned}
J_1 &= \frac{1}{6(U+J_H)}(5t^2 -6\lambda_{1y}^2 - 8\sqrt{2}\lambda_{1z}\lambda_{1y} - 6\lambda_{1z}^2) \\
K_1 &= \frac{2}{U+J_H}\lambda_{1y}(\lambda_{1y}+2\sqrt{2}\lambda_{1z}) \\
\Gamma_1 &= \frac{2}{3(U+J_H)} (\lambda_{1y} - \sqrt{2}\lambda_{1z})^2 \\
\Gamma_1' &= -\frac{4}{3(U+J_H)} (\lambda_{1y} - \sqrt{2}\lambda_{1z})(\lambda_{1y}+\frac{1}{\sqrt{2}}\lambda_{1z}) ,\\
J_3 &= \frac{t_3^2}{U+J_H} ,
\end{aligned}
\end{gather}
where $t_1^2\equiv t_{1a}^2 + t_{1b}^2$ and $t_3^2\equiv t_{3a}^2 + t_{3b}^2$. 
The effective interaction $U+J_H$ can be obtained using published results for the
constrained RPA (cRPA) calculations of $U,J_H$ \cite{yekta2021strength}; 
this effective coupling, listed in Table.\ref{NiX2 exchange parameters}, consistently 
decreases from Cl to Br to I, which can be understood in terms of increasing $d$-$p$ hybridization along Cl-Br-I series, which
enhances the screening.

Using the estimates of $t_{1\alpha}, t_{3\alpha}, \lambda_{1z}, \lambda_{1y}$ for the effective two-orbital model (see Table I) 
for Ni$X_2$, we computed the exchanges,
$J_1$, $K_1$, $\Gamma_1$, $\Gamma_1'$ and $J_3$, which are listed in Table.\ref{NiX2 exchange parameters} and compared with DFT estimated exchanges from
Ref.\cite{riedl2022microscopic}.
As seen from Table \ref{NiX2 exchange parameters}, we find reasonably good agreement between
the $e_g$-orbital model results and previous published DFT values \cite{riedl2022microscopic} for most
exchange couplings for different materials.
The single major discrepancy is in the 1NN Heisenberg term $J_1$ which explicitly needs the ligand site;
a previous calculation of this FM Goodenough-Kanamori exchange on the 90${}^\circ$ $M$-$X$-$M$ bond
within a $d$-$p$ model and incorporating Hubbard repulsion and Hund's coupling
on the ligand site 
was shown to capture results from the DFT four-state calculation 
\cite{riedl2022microscopic}. 

We note that our results for the bond-anisotropic exchange
couplings are comparable to those extracted from numerical exact diagonalization using the full 
five-orbital model \cite{riedl2022microscopic};
the small differences between our results can be attributed to $t_{2g}$-$e_g$ hopping terms in the five-orbital model
and to differences in the effective Coulomb interactions.
The advantage of our work over Ref.\cite{riedl2022microscopic} is that our approach provides an
analytical expression for the exchange couplings, which helps in
microscopic insights on the dependency of material-specific parameters.

 \begin{table}[t]
\caption{Exchange parameters (in meV) for Ni$X_2$ obtained from the effective two-orbital model using parameters from Table I.
We use the indicated effective $U+J_H$ for each material from
Yekta et al \cite{yekta2021strength}. For comparison, we show results for the exchange couplings from four-state
DFT calculations \cite{riedl2022microscopic}.
Aside from the nearest-neighbor Heisenberg FM coupling $J_1$, which needs explicit ligand mediated Goodenough-Kanamori
exchange on the $90^\circ$ $M$-$X$-$M$ bond, all other exchange couplings are reasonably reproduced by the two-orbital model.}
% Instead, the perturbative result in the $dp$ model, denoted 
% $J_1^{dp}$, may be compared with DFT results.}
\label{NiX2 exchange parameters}
\begin{ruledtabular}
\begin{tabular}{lccc@{\qquad}ccc}
 & \multicolumn{3}{c}{Effective $e_g$ model} & \multicolumn{3}{c}{Ref.~\cite{riedl2022microscopic}} \\
\cline{2-4}\cline{5-7}
Ni$X_2$ & NiCl$_2$ & NiBr$_2$ & NiI$_2$ & NiCl$_2$ & NiBr$_2$ & NiI$_2$ \\
$(U+J_H)$  &$3.84\mathrm{eV}$ &$3.5\mathrm{eV}$ &$2.65\mathrm{eV}$ \\
\hline
$J_1$         & $\approx 0.0$ & $0.01$ & $-0.09$ &   $-2.9$ & $-3.9$ & $-6.2$  \\
$K_1$         & $+0.01$  & $+0.17$ & $+1.43$ & $\approx 0.0$  & $+0.2$ & $+2.2$ \\
$\Gamma_1$    & $\approx 0.0$  & $\approx 0.0$  & $0.02$  & $\approx 0.0$  & $\approx 0.0$  & $\approx 0.0$ \\
$\Gamma_1'$   & $\approx 0.0$  & $\approx 0.0$  & $-0.16$ & $\approx 0.0$  & $\approx 0.0$  & $-0.1$ \\
$J_3$         & $+0.91$ & $+1.42$ & $+3.64$ & $+0.8$ & $+1.8$ & $+4.2$
% $J_1^{dp}$         & $-1.6$ & $-1.5$ & $-1.9$ & $-2.9$ & $-3.9$ & $-6.2$
\end{tabular}
\end{ruledtabular}
\end{table}

% $\Delta J_1 = -\frac{1}{2} \frac{t_{pd\sigma}^4}{\Delta_{CT}^2} \big( \frac{1}{E_T} - \frac{1}{E_S} \big)$. 
% Here $E_{T} = 2\Delta_{CT}+U_p - 3J_H^p$ and $E_{S} = 2\Delta_{CT}+U_p - J_H^p$ denote the energies of the two-hole triplet/singlet 
% states on the ligand site, with $U_p$ and $J_H^p$ the Hubbard interaction and Hund's coupling on the p-orbitals, respectively. Here $\Delta_{CT} = E(d^9p^5) - E(d^8p^6) = \Delta_{pd}+ U+V+J_H^d/2 =\Delta_{pd} + 2U_d - 2J_H^d$ \textcolor{red}{Check? Why $2 U_d-2 J_H$?} is the charge transfer gap, with $U_d$ and $J_H^p$ the Hubbard interaction and Hund's coupling on the Ni-$3d$ orbitals. 

% From Table \ref{NiX2 exchange parameters}, we see that the prior results in Ref.~\cite{riedl2022microscopic}
% show that the FM 1NN Heisenberg interaction $J_1$ is
% the most dominant one, followed by AFM Heisenberg 3NN interaction $J_3$ and subdominant Kitaev interaction, 
% $K_1$, while the anisotropic $\Gamma_1$ and $\Gamma_1'$ terms are negligibly small. From our results,
% for the effective $e_g$ model,
% we also find negligible $\Gamma_1$ and $\Gamma_1'$, and values for $J_3, K_1$ which are in reasonable
% agreement with Ref.~\cite{riedl2022microscopic}. Going beyond the $e_g$ model to compute $J_1^{dp}$ using
% perturbation theory, we find results somewhat smaller than
% previously reported values \cite{riedl2022microscopic}, likely reflecting differences in the DFT estimates 
% of the parameters used in our calculation and limitations of the perturbative approach to the $dp$ model (see SM \cite{suppmat}).

\begin{figure}[h]
\centering
\includegraphics[width=0.48\textwidth]{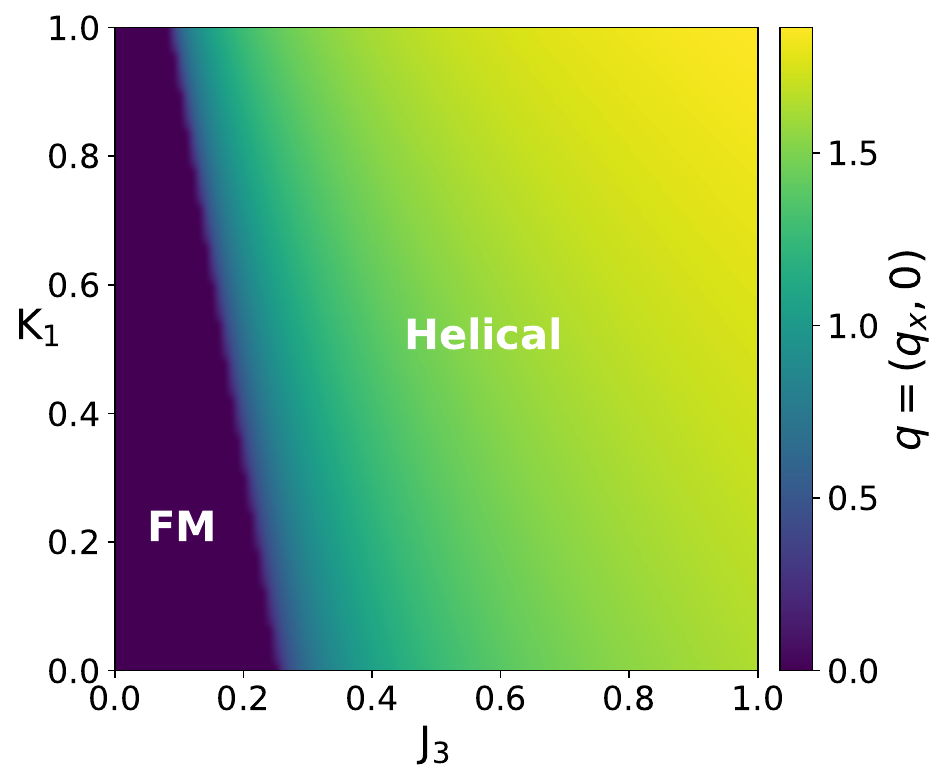}
\caption{Luttinger-Tisza (LT) phase diagram of the triangular lattice $J_1$-$J_3$-$K_1$ model, with $J_1 = -1$. The ordered states are characterized by a wave vector $(q_x, 0)$. Ferromagnetic (FM) and spin helical phases have been labeled. The transition between the
FM and Helical happens for $K_1 = -\frac{3}{2}J_1 - 6J_3$. The LT solutions identify the ordering wave vector, although the corresponding eigenmodes do not generally satisfy the local fixed spin-length constraint for $K_1 \neq 0$ in the helical phase.}
\label{LT J1-J3-K phase diagram}
\end{figure}

For completeness, Fig.~\ref{LT J1-J3-K phase diagram} shows
the magnetic phase diagram of the $J_1$-$J_3$-$K_1$ model obtained using the 
standard Luttinger-Tisza method \cite{LTmethod}, with the ferromagnetic exchange fixed to $J_1= -1$. The phase diagram hosts collinear
ferromagnetic and incommensurate helical orders, both characterized by ordering wave vectors $\vec{q} = (q_x, 0)$ along the $\Gamma-K$ line. 
This model-level phase diagram is broadly in conformity with the magnetic order in a broad class of {1T-$MX_2$} materials, which were first established 
in bulk crystals\cite{kuindersma1981magnetic} and have recently been reported in the exfoliated
monolayer $\mathrm{NiI_2}$ \cite{song2022evidence}. 
Several first-principles calculations have also proposed related spin models for these materials \cite{ni2021giant, riedl2022microscopic, li2023realistic}.
These supports the ferromagnetic ground state in $\mathrm{NiCl_2}$ \cite{lines1963magnetic}, and the helical ground states in $\mathrm{NiBr_2}$ and $\mathrm{NiI_2}$ \cite{mcguire2017crystal}, consistent with our findings.

% The phase diagram (cf Fig.\ref{LT J1-J3-K phase diagram}) in $J_3$-$K_1$ parameter space shows the collinear FM to incommensurate spiral ordering to commensurate striped order.
% \textcolor{red}{ADD MORE}

\subsection{Polarization and Spin currents}
The noncolinear helical phases discussed above are directly relevant to type-II multiferroicity that has been experimentally realized in a broad class of {1T-$MX_2$} materials \cite{tokunaga2011multiferroicity, kurumaji2011magnetic, kurumaji2013magnetoelectric, cahlik2026universality}.
This response is commonly described using the DFT-based generalized spin-current method \cite{xiang2011general}, which has been widely applied to  $\mathrm{MnI_2}$ \cite{xiang2011general}, $\mathrm{VX_2 (X = Cl, Br, I)}$ \cite{liu2024spin}, and $\mathrm{NiX_2 (X = Cl, Br, I)}$ \cite{pan2025long, yu2025microscopic, ghojavand2026strain}. Within this framework, the coupling between the electric polarization and the spins on a given bond $(i,j)$ is expressed as 
\begin{gather}
\vec{P}_{ij} = \vec{M}_{ij} (\vec{S}_i \times \vec{S}_j). 
\end{gather}
where $\vec{M}_{ij}$ is the rank-2 magneto-electric tensor.
The microscopic origin of this inverse Dzyaloshinskii-Moriya effect and its connection to spin currents
was first proposed by \cite{katsura2005spin}, and was further studied in spin-orbit-coupled $t_{2g}$ Mott insulators \cite{bolens2018theory, solovyev2021magnetically}. 
In particular, Ref.~\cite{solovyev2021magnetically} developed a strong-coupling formulation for a one-orbital $j_{\text{eff}} = 1/2$ Mott insulator, in which the bond polarization is generated through virtual hole-doublon processes. Within this one-orbital setting, the combination of the pseudospin-independent hopping $t_{ij}^0$ and the pseudospin-dependent projected position matrix elements $\vec{r}_{ij}$ can produce a finite spin-induced polarization even on a centrosymmetric bond. \\

% \textcolor{magenta}{TSD: Parts of the following paragraph should go after the general part!}
In the present work, we extend this perspective to the two-orbital $e_g$ model, as relevant to 1T-$MX_2$ materials, where the effective SOC originates from the ligand $p$-orbitals. After projection to the low-energy $e_g^*$ subspace, the ligand SOC generates both spin-dependent hopping processes and spin-dependent matrix elements of the position operator. 
We start by deriving the most general form of the spin-induced polarization for a two orbital model. The local Hermitian spin operators are
\begin{gather}
\begin{aligned}
&\vec{S}_{ia} = a^\dag_i \bm{\sigma} a^\pdg_i, \quad \vec{S}_{ib} = b^\dag_i \bm{\sigma} b^\pdg_i,\\ 
&\vec{S}_{ic} = a^\dag_i \bm{\sigma} b^\pdg_i + b^\dg_i \bm{\sigma} a^\pdg_i, \quad \vec{S}_{is} = -i(a^\dag_i \bm{\sigma} b^\pdg_i 
- b^\dg_i \bm{\sigma} a^\pdg_i) .
\end{aligned}
\end{gather}
Here $\vec{S}_{ia}$ and $\vec{S}_{ib}$ are the spin operators for orbitals $a$ and $b$ respectively, while  
$\vec{S}_{ic}, \vec{S}_{is}$ describe orbital-mixing spin degrees of freedom, and the site index is denoted by $i$. Among these operators, $S_{iA} (A \in \{a,b,c\})$ are time-reversal odd, while $\vec{S}_{is}$ is time-reversal even. Therefore $\vec{S}_{is}$ does not couple to other spin operators in the time-reversal-even bilinear terms in the bond polarization expression.
We consider orbitals with the same inversion parity; for $e_g$ manifold both orbitals are even parity. 
To quadratic order in spin operators, the spin-induced polarization $\vec{P}_{ij}$ on a centrosymmetric bond $(ij)$ 
takes the general form
\begin{gather}
\begin{aligned}
\!\!\!\! \vec{P}_{ij} = &\!\!\!\!\! \sum_{\mu \in \{a,b,c,s\}} \!\!\!\!\!\! \underline{\vec{D}}_{ij}^{\mu\mu} (\vec{S}_{i\mu} \times \vec{S}_{j\mu})\\
&+\!\!\!\!\! \sum_{\mu > \nu \in \{a,b,c\}} \!\!\!\!\!\! \underline{\vec{D}}_{ij}^{\mu \nu} \cdot (\vec{S}_{i\mu} \times \vec{S}_{j\nu} + \vec{S}_{i\nu} \times \vec{S}_{j\mu}) \\
&+\!\!\!\!\! \sum_{\mu > \nu\in \{a,b,c\}} \!\!\!\!\!\! (\vec{S}_{i\mu}\cdot \overleftrightarrow{\Gamma}_{ij}^{\mu\nu} \cdot \vec{S}_{j\nu} - \vec{S}_{i\nu}\cdot \overleftrightarrow{\Gamma}_{ij}^{\mu\nu} \cdot \vec{S}_{j\mu}),
\end{aligned}
\end{gather}
where $\mu=a,b,c,s$, the rank-2 tensor $\underline{\vec{D}}_{ij}^{\mu \nu}$ leads to a spin-antisymmetric contribution, and the rank-3 tensor $\overleftrightarrow{\Gamma}^{\mu\nu}_{ij} = [\big(\Gamma^{\mu\nu}_{ij}\big)^x, \big(\Gamma^{\mu\nu}_{ij}\big)^y, \big(\Gamma^{\mu\nu}_{ij}\big)^z] \, (\mu \neq \nu)$ corresponds to spin-symmetric contribution.

Additional bond symmetries further constrain the couplings. In particular, for the $\delta_1$-bond with $C_{2x}$ symmetry, the allowed matrix structures are given by
\begin{gather}
\begin{aligned}
\underline{\vec D}^{\mu \nu} =  \begin{bmatrix} D_{xx} &0 &0\\ 0 &D_{yy} &D_{yz} \\ 0 &D_{zy} &D_{zz} \end{bmatrix}^{\mu \nu}, 
\big(\Gamma^{\mu \nu}\big)^x =  \begin{bmatrix} \Gamma_{xx}^{x} &0 &0\\ 0 &\Gamma_{yy}^{x} &\Gamma_{yz}^{x} \\ 0 &\Gamma_{yz}^{x} &\Gamma_{zz}^{x} \end{bmatrix}^{\mu\nu}, \\[5pt]
\big(\Gamma^{\mu \nu}\big)^y =  \begin{bmatrix} 0 &\Gamma_{xy}^y &\Gamma_{xz}^y\\ \Gamma_{xy}^y &0 &0 \\ \Gamma_{xz}^y &0 &0 \end{bmatrix}^{\mu \nu}, \quad 
\big(\Gamma^{\mu \nu}\big)^z =  \begin{bmatrix} 0 &\Gamma_{xy}^z &\Gamma_{xz}^z\\ \Gamma_{xy}^z &0 &0 \\ \Gamma_{xz}^z &0 &0 \end{bmatrix}^{\mu \nu}.
\end{aligned}
\end{gather}

In contrast to the single-orbital case \cite{katsura2005spin,solovyev2021magnetically},
the polarization in our two-orbital model has terms proportional to $\underline{\vec{D}}$ which depend 
on intra-orbital spin cross-products,
such as  $\vec{S}_{ia} \times \vec{S}_{ja}$ or $\vec{S}_{ib} \times \vec{S}_{jb}$, as well as 
inter-orbital spin cross-products, such as 
$\vec{S}_{ia} \times \vec{S}_{jb}$. These will be nonzero for non-colinear spin arrangements.
As in the seminal work of KNB \cite{katsura2005spin}, we can relate the spin cross-products appearing with $\vec{D}$
to ground state spin currents of the electrons which for our model includes both intra-orbital spin currents
and inter-orbital spin currents.
% generalized spin-current operators, such as intra-orbital spin currents
% which are proportional, for instance, to $\vec{S}_{ia} \times \vec{S}_{ja}$ or $\vec{S}_{ib} \times \vec{S}_{jb}$, 
% as well as inter-orbital spin currents which arise from
% $\vec{S}_{ia} \times \vec{S}_{jb}$. In addition, there are distinct polarization terms embedded in 
% $\overleftrightarrow{\Gamma}$, which includes
% operators such as $(\vec{S}_{ia}\cdot\vec{S}_{jb} - \vec{S}_{ib}\cdot \vec{S}_{ja})$ which are unrelated to spin
% currents and specific to multiorbital systems. 
Table.~\ref{fermionic operators and spin bilinear} lists the multi-orbital
spin current operators together with their 
corresponding bilinear spin operators after projection to the local spin degree of freedom.
\begin{table}[h]
\caption{Correspondence between fermionic spin-current operators and bilinear spin operators for
two-orbital $e_g$ model.}
\label{fermionic operators and spin bilinear}
\centering
\begin{tabular}{cc}
\hline\hline
Fermion & Spin \\
\hline
$\sim i(a_i^\dag \Vector{\sigma} a_j - a_j^\dag \Vector{\sigma} a_i)$ &\quad $\vec{S}_{i\mu} \times \vec{S}_{j\mu} \, (\mu = a,c,s)$ \\
$\sim i(b_i^\dag \Vector{\sigma} b_j - b_j^\dag \Vector{\sigma} b_i)$ &\quad  $\vec{S}_{i\mu} \times \vec{S}_{j\mu} \, (\mu = b,c,s)$ \\
$\sim i(a_i^\dag \Vector{\sigma}b_j + b_i^\dag \Vector{\sigma}a_j - \text{h.c.} )$ &\quad $\vec{S}_{i\mu}\times \vec{S}_{jc} + \vec{S}_{ic}\times \vec{S}_{j\mu} \, (\mu = a,b)$  \\
\hline\hline
\end{tabular}
\end{table}

In addition, there are distinct polarization terms embedded in 
$\overleftrightarrow{\Gamma}$ which are specific to the two-orbital scenario but which trivially
vanish for the single-orbital example. Such
operators include $(\vec{S}_{ia}\cdot\vec{S}_{jb} - \vec{S}_{ib}\cdot \vec{S}_{ja})$ which 
can potentially contribute to the polarization even for {\it collinear} orbital-dependent spin orders.

% \textcolor{red}{Expand on this in few sentences about connection to spin current language as extension of KNB work}

Deep in the Mott insulator, Hund's coupling locks the two $e_g$ electrons into a local $S = 1$ moment, so the orbital degree of
freedom gets quenched, with $\vec{S}^a = \vec{S}^b$. Furthermore, we expect $\vec{S}^c \approx 0, \vec{S}^s \approx 0$;
since a nonzero value of these require mixing with doublon states which would be suppressed in the Mott insulator.
In this case, only the spin-antisymmetric parts survive, and the polarization then reduces to the conventional 
generalized spin-current form,
\begin{gather}
\vec{P}_{ij} = \underline{\vec{D}}_{ij}^{\text{eff}} \big(\vec{S}_{i} \times \vec{S}_{j}\big),
\end{gather}
where $\vec{S}_i$ is the effective $S = 1$ spin operator. The effective coupling $\underline{\vec{D}}_{ij}^{\text{eff}} = (\underline{\vec{D}}_{ij}^{aa} + \underline{\vec{D}}_{ij}^{bb} + 2 \underline{\vec{D}}_{ij}^{ab})/4$, where $\vec{S}_{ia} = \vec{S}_{ib} = \vec{S}_{i}/2$ in the local triplet manifold.
Despite this similarity, the resultant magnetoelectric coupling reflects the two-orbital structure: 
orbital-off-diagonal position matrix elements can combine with SOC-induced orbital-off-diagonal hoppings and 
contribute to the bond polarization shown below.

% \textcolor{magenta}{In the deep Mott limit, as applicable for Ni halides, Nevetheless,  Hund's coupling locks the two $e_g^*$ spins into a local $S = 1$ moment, so that} 
% %we focus on the local $S = 1$ triplet sector and take 
% $\vec{S}^a = \vec{S}^b$ and $\vec{S}^c = \vec{S}^s = 0$. \textcolor{magenta}{Further in the Mott limit, nonzero $\vec{S}_{ic}$ and $\vec{S}_{is}$ require mixing with doublon states and are therefore suppressed.}
% %deviations from $\vec{S}_{ia} = \vec{S}_{ib}$ are suppressed by Hund's coupling, while nonzero $\vec{S}_{ic}$ and $\vec{S}_{is}$ require mixing with doublon states and are therefore suppressed.
% In this case, only the spin-antisymmetric parts survive. The polarization then reduces to the conventional generalized spin-current form,
% \begin{gather}
% \vec{P}_{ij} = \vec{D}_{ij}^{\text{eff}} \big(\vec{S}_{i} \times \vec{S}_{j}\big),
% \end{gather}
% where $\vec{S}_i$ denotes the effective local $S = 1$ spin operator. The effective coupling is given by $\vec{D}_{ij}^{\text{eff}} = (\vec{D}_{ij}^{aa} + \vec{D}_{ij}^{bb} + 2 \vec{D}_{ij}^{ab})/4$, where $\vec{S}_{ia} = \vec{S}_{ib} = \vec{S}_{i}/2$ in the local triplet manifold.

In the Mott limit, the polarization can be calculated by treating the hopping $\hat{T}$ as perturbation \cite{solovyev2021magnetically}, starting from the $S = 1$ ground states on each site, $|g_i\rangle = |\chi_i\rangle_a |\chi_i\rangle_b$ where $|\chi_i\rangle = e^{-i\phi_i}\cos \frac{\theta_i}{2} |\ua\rangle + \sin \frac{\theta_i}{2} |\da\rangle$. The bond polarization is given by
\begin{gather}\label{bond polarization expression}
\Vector{P}_{ij} = \xi \langle g_i g_j| \hat{\Vector{r}}_{(ij)} \hat{T}_{(ij)} + \hat{T}_{(ij)} \hat{\Vector{r}}_{(ij)} |g_i g_j\rangle ,
\end{gather}
where $\xi \equiv e/(U+J_H)$, $\hat{T}_{(ij)}$ is the hopping Hamiltonian on the bond $(ij)$, and the bond position operator is
\begin{gather}
\hat{\Vector{r}}_{(ij)} = \Psi_i^\dag \Vector{r}_{ij} \Psi_j + \Psi_j^\dag \Vector{r}_{ij}^\dag \Psi_i .
\end{gather}
Inversion symmetry and Hermiticity of $\hat{\Vector{r}}_{(ij)}$ requires the matrices to be anti-Hermitian, i.e, $\Vector{r}_{ij}^\dag = -\Vector{r}_{ij}$. Focusing on the 1NN $\delta_1$-bond and the 3NN $2\delta_1$-bond, the $C_{2x}$ bond symmetry further constrains the allowed terms to take the form
\begin{align}
r_{ij}^x &= i\begin{bmatrix} R_{ax}^x &0 \\ 0 &R_{bx}^x \end{bmatrix} \sigma_x + R^x_{xy} i\tau_x \sigma_y+ R^x_{xz} i\tau_x \sigma_z, \\
r_{ij}^{\alpha} &= R_{y0}^{\alpha} i\tau_y \sigma_0 + i\begin{bmatrix} R_{ay}^{\alpha} &0 \\ 0 &R_{by}^{\alpha} \end{bmatrix} \sigma_y + i\begin{bmatrix} R_{az}^{\alpha} &0 \\ 0 &R_{bz}^{\alpha} \end{bmatrix} \sigma_z \notag \\
&\quad + R_{xx}^{\alpha} i\tau_x \sigma_x \quad (\alpha = y,z),
\end{align}
where $\tau_{0,x,y,z}$ and $\sigma_{0,x,y,z}$ are Pauli matrices in the orbital and spin spaces, respectively. The coefficients $R_{\alpha \beta}^{\gamma}$ contain both spin-dependent and spin-independent contributions to the projected position operator. The spin-dependent components arise because the effective $e_g^*$ states contain ligand-$p$ components, which are further mixed by ligand SOC. In the SM Section V \cite{suppmat}, we use perturbation theory to estimate this contribution.
For both 1NN and 3NN bonds, the leading spin-dependent contribution scales as $\sim \lambda_p t_{pd\sigma}/{\Delta_*^2}$, which is linear in $\lambda_p$ in the regime $\lambda_p \ll \Delta_*$, as are the spin-dependent hoppings. Evaluating Eq.~\eqref{bond polarization expression}, we obtain the following magnetoelectric couplings for the 1NN $\delta_1$-bond:
\begin{gather}\label{SE eqns for polarization}
\begin{aligned}
\!\!\! \underline{\vec D}_{xx}^{\text{1NN}} &= 2 \xi \big( t_{1a} R_{ax}^{x,\text{1NN}} + t_{1b} R_{bx}^{x,\text{1NN}} \big), \\
\!\!\! \underline{\vec D}_{yy}^{\text{1NN}} &= 2 \xi \big( t_{1a} R_{ay}^{y,\text{1NN}} + t_{1b} R_{by}^{y,\text{1NN}} + 2\lambda_{1y} R_{y0}^{y,\text{1NN}} \big), \\
\!\!\! \underline{\vec D}_{yz}^{\text{1NN}} &= 2 \xi \big( t_{1a} R_{az}^{y,\text{1NN}} + t_{1b} R_{bz}^{y,\text{1NN}} - 2\lambda_{1z} R_{y0}^{y,\text{1NN}} \big), \\
\!\!\! \underline{\vec D}_{zy}^{\text{1NN}} &= 2 \xi \big( t_{1a} R_{ay}^{z,\text{1NN}} + t_{1b} R_{by}^{z,\text{1NN}} + 2\lambda_{1y} R_{y0}^{z,\text{1NN}} \big), \\
\!\!\! \underline{\vec D}_{zz}^{\text{1NN}} &= 2 \xi \big( t_{1a} R_{az}^{z,\text{1NN}} + t_{1b} R_{bz}^{z,\text{1NN}} - 2\lambda_{1z} R_{y0}^{z,\text{1NN}} \big).
\end{aligned}
\end{gather}
Similar results can be obtained for 2NN and 3NN bonds. Eq.~\ref{SE eqns for polarization} constitutes one of the key results of
our work.

To numerically estimate $R_{\alpha \beta}^{\gamma}$, we use hydrogen-like atomic orbital wavefunctions with effective nuclear charges $Z^{\text{eff}}_{\mathrm{Ni, 3d}} = 12.53$ and $Z^{\text{eff}}_{\mathrm{I, 5p}} = 11.61$ \cite{clementi1963atomic, clementi1967atomic}, together with perturbative corrections to the wave functions using a local $\mathrm{NiX_6}$ cluster. The resulting values for
$R_{\alpha \beta}^{\gamma}$ are listed in SM Section V.D \cite{suppmat}. We find $(R_{\alpha \beta}^{\gamma})_{\text{1NN}} \sim 10^{-2}\AA$ for 1NN bond, and $(R_{\alpha \beta}^{\gamma})_{\text{3NN}} \sim 10^{-3}\AA$ for 3NN bond. Using Eq.~\eqref{SE eqns for polarization}, we obtain the following magnetoelectric matrices  (in units of $10^{-5} e\AA$),
\begin{gather}
\underline{\vec D}^{\text{1NN}} = \begin{bmatrix} 0 &0 &0\\ 0 &82 &-60 \\ 0 &126 &-88  \end{bmatrix}, \quad \underline{\vec D}^{\text{3NN}} = \begin{bmatrix} 0 &0 &0\\ 0 &-17 &4 \\ 0 &17 &19  \end{bmatrix} .
\end{gather}
while $\underline{\vec D}^{\text{2NN}}$ is found to be negligible.
% These matrices show substantial deviations from the KNB expression for both 1NN and 3NN magnetoelectric couplings, \textcolor{magenta}{TSD: In what sense?} 
These results are qualitatively consistent with previous first-principles calculations \cite{pan2025long, yu2025microscopic}, having right
order of magnitudes. Our estimates also capture  the hierarchy of couplings: $\underline{\vec D}^{\text{3NN}}$ is comparable to 
$\underline{\vec D}^{\text{1NN}}$, owing to the dominant $t_{3b}$ hopping and the extended ligand weight of the effective orbitals, whereas $\underline{\vec D}^{\text{2NN}}$ is strongly suppressed by its much smaller hopping amplitudes. There
are though some quantitative differences compared to DFT results. For example, $\underline{\vec D}_{xx}$ turns out to be nonzero
in DFT results \cite{pan2025long, yu2025microscopic}, while it is zero in our result. Similarly,
the DFT values of $(\underline{\vec D}^{\text{1NN}})_{zy}$ and $(\underline{\vec D}^{\text{1NN}})_{zz}$ are found to much smaller
compared to our result, while $\underline{\vec D}^{\text{3NN}}$ is larger in DFT. These quantitative discrepancies arise due to
approximations concerning atomic wavefunction, the perturbative treatment of ligand SOC within an isolated $\mathrm{NiX_6}$ cluster that neglects long-range
ligand-ligand hopping effects.
%the overall order of magnitude, it does not quantitatively reproduce all matrix elements. For example, the DFT values of $(D^{\text{1NN}})_{zy}$ and $(D^{\text{1NN}})_{zz}$ are nearly zero, while the DFT $D^{\text{3NN}}$ is larger in DFT. This discrepancy is expected, since the hydrogenlike atomic estimation is only approximate, and the perturbative treatment of ligand SOC within an isolated $\mathrm{NiX_6}$ cluster neglects long-range ligand-ligand hopping effects.  Nevertheless, the symmetry analysis remains applicable to the Wannier-orbital representation of the low-energy bands. This explains \\

From our ``super-exchange'' derivation in Eq.~\eqref{SE eqns for polarization}, together with perturbative estimates of the spin-dependent matrix elements $R_{\alpha \beta}^\gamma$ and spin-dependent hoppings $(\lambda_{1y}, \lambda_{1z}, \lambda_{3y}, \lambda_{3z})$, we expect both $\underline{\vec D}^{\text{1NN}}$ and $\underline{\vec D}^{\text{3NN}}$ scale linearly with $\lambda_p$ in the regime ligand SOC $\lambda_p \ll \Delta_*$, which aligns with previous DFT results \cite{yu2025microscopic, pan2025long} and minimal cluster models \cite{pan2025long, jia2007microscopic}.

\section{Summary}
In this work, we have shown that an 
effective two-orbital $e_g$ model with SOC-dependent hopping terms displays 
rich physics of band topology and correlation driven phenomena. In particular,
in the non-interacting or weak correlation limit, the presence of effective spin-preserving and spin-mixing hopping terms, arising from the SOC at the downfolded ligand site, result in non-trivial topological phases. While turning on the spin-preserving SOC hopping term, makes the bands to be characterized by non-zero even integer values of spin-Chern index and dissipation-less metallic edge-state spectra, inclusion of the additional spin-mixing terms stabilizes higher-order topology characterized by corner states and a fractional corner anomaly. In the strongly interacting Mott limit, the $S=1$ spin model shows interplay of
Heisenberg and Kitaev interactions, stabilizing collinear ferromagnetic or helical spin order. Finally, for intermediate to strong
correlations, we have derived an extension of the magneto-electric effect beyond the single-orbital case where multi-orbital nature of electronic 
levels can interplay with orbital-dependent spin textures to yield new terms in the expression for the polarization. The implications of this generalized magneto-electric effect to multi-orbital insulators and its connection to multipolar
currents beyond the spin-current framework will be taken up in a follow-up work.
% gives rise to
% the magneto-electric effect which also shows additional contributions due to the multi-orbital nature of the problem, compared to single-orbital scenario.

The material realizations of the emergent phenomena discovered in our model deserve further study. For example, to stabilize 
the non-trivial spin-Chern or HOT,
suppression of magnetic order is necessary. Similarly,
to uncover the multi-orbital fingerprints in magnetoelectric effect, the compounds need to be pushed away 
from the deep Mott insulator regime to be closer to metal-insulator boundary. Such tuning of material properties
may conceivably be achieved using strain or pressure or doping.

\acknowledgments
We acknowledge helpful discussions with Mark Hirsbrunner and Riccardo Comin.
AP and TSD acknowledge financial support from the Anusandhan National Research Foundation (ANRF),
Government of India, under the collaborative VAIBHAV Programme [Grant No.INAE/DST/VF/2024/II/01].
AP acknowledges funding from
the Natural Sciences and Engineering Research Council of Canada (NSERC) 
through Discovery grant RGPIN-2026-04578.
YX acknowledges support from the NSERC through a Canada Graduate Research Scholarship (CGRS-D).
\bibliography{citation}

\clearpage
\onecolumngrid
\begin{center}
{\Large \bf Supplemental Material}
\end{center}

\beginsupplement

% \appendix

\section{Density functional theory results}
First-principles DFT calculations were performed using a plane-wave basis and projector augmented-wave (PAW) potentials~\cite{PhysRevLett.77.3865}, as implemented in the Vienna \textit{Ab initio} Simulation Package (VASP)~\cite{PhysRevB.50.17953,TACKETT2001348,10.1063/1.1926272}. The exchange--correlation functional was treated within the Perdew--Burke--Ernzerhof (PBE) generalized gradient approximation (GGA)~\cite{PhysRevLett.77.3865}. Convergence of the total energies and forces was ensured by employing a plane-wave energy cutoff of 600~eV and a $6\times 6 \times 1$ Monkhorst--Pack grid for Brillouin-zone sampling. During the structural relaxations, all ionic positions were optimized until residual forces on each atom were reduced below 0.0001~eV/\AA.
\begin{figure}[h]
    \centering
    \includegraphics[width=0.6\linewidth]{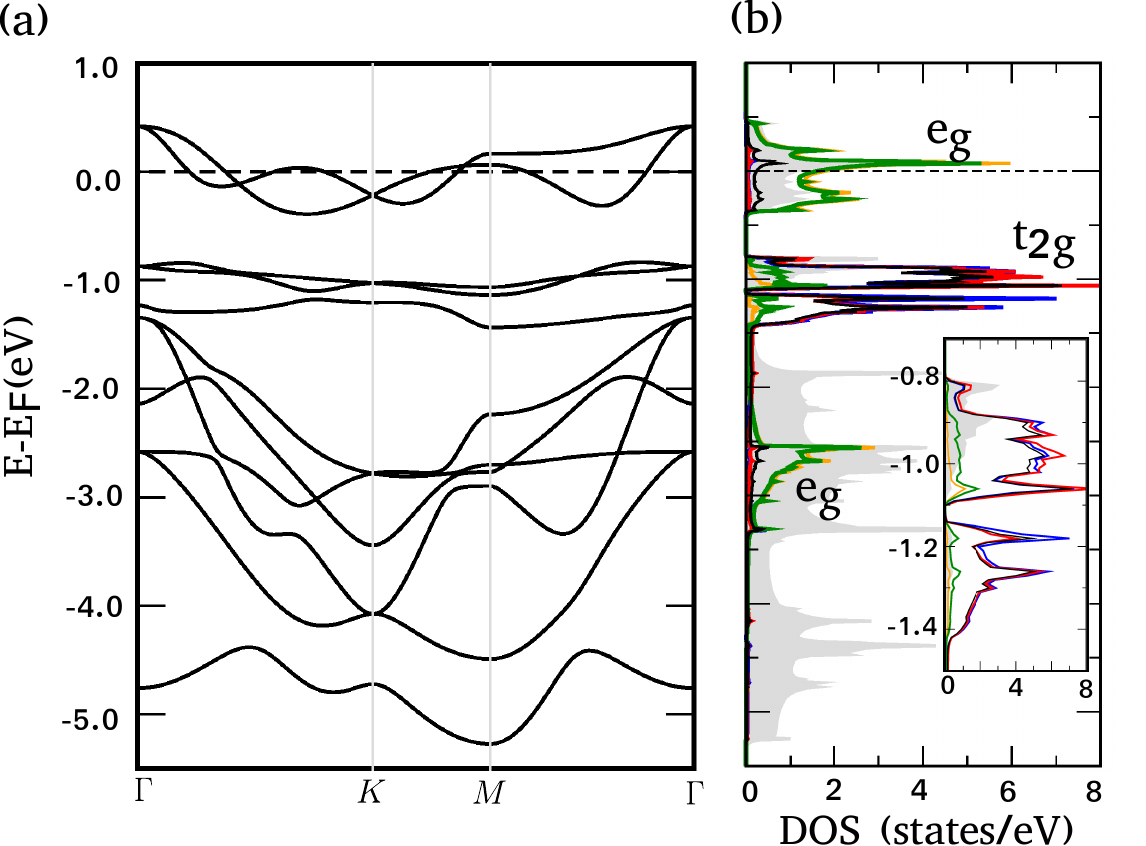}
    \caption{The non-spin-polarized GGA density of states and band-structure of NiI$_2$. (a) The 5+6 bands for one formula unit of NiI$_2$, (b) Density of states, projected to different Ni-$d$ characters, $d_{XY}$ (black), $d_{YZ}$ (red), $d_{XZ}$ (blue), $d_{3Z^{2}-r^2}$ (green) and $d_{X^2-Y^2}$ (orange) and I-$p$ (filled gray area). Zero of the energy is set at E$_F$. An inset is included for the $t_{2g}$-DOS for better visibility.}
    \label{fig:DFT_bands}
\end{figure}
The non-spin-polarized GGA-PBE calculations for the band-structure and density of states reveal the electronic characteristic for NiI$_{2}$ as presented in Fig.~\ref{fig:DFT_bands}.  We have 11 bands (c.f. Fig~\ref{fig:DFT_bands}(a)), contributed by 1 Ni (five d-orbitals) and 2 I (six p-orbitals) ions, spanning the energy range $\approx$-5.5eV to 0.5 eV. Fig~\ref{fig:DFT_bands}(b) presents the density of states (DOS) of NiI$_2$, projected onto Ni-$d$ and I-$p$, with choice of  $X$, $Y$ and $Z$ in local coordinate system as shown in Fig~2 of the main text.  
We have two $e_{g}$ bands lying in the energy range $\approx$ -0.5 to 0.5 eV. However, $e_{g}$ orbitals are highly hybridized with the I-$p$ orbitals leading to orbital character distribution, due to bonding and anti-bonding type of interaction, among I-p and Ni-$e_{g}$ bands. Due to this, we have significant I-$p$ DOS in the range $\approx$ -0.5 eV to 0.5 eV and Ni-$e_{g}$ character in the energy range $\approx$ -3.5 eV to -2.2 eV.  The $t_{2g}$ orbitals are broadly restricted in energy range from $\approx$ -1.5 eV to -0.75 eV with small hybridization with the I-$p$ orbitals.
Since the low energy $t_{2g}$ orbitals are fully occupied, we develop the 
minimal two-orbital model for Ni$X_2$ using only the Ni-$e_g$ orbitals, 
incorporating the hybridization with ligands in terms of renormalized parameters and additionally including an effective SOC. 

Our results for the $dp$ model
from Wannierization 
are summarized in Table \ref{dp_model DFT parameters} below, along with the ligand SOC inferred from 
spectroscopic data \cite{dagenais1976precise, arnold1995study}.
For the effective two-orbital model, the extracted 
hopping parameters from downfolding
to the effective $e_g$ orbitals are listed in Table I of the main text.
% \begin{table}[h]
% \begin{centering}
% \begin{tabular}{|l|l|l|l|l|l|}
% \hline
%  Parameters (eV) &~$t_{pd\sigma}~$ &~$t_{pp\sigma}$~  &~$t_{pp\pi}/t_{pp\sigma}$~  &~$\Delta_{pd}$~ &$\lambda_p$ \\ \hhline{|=|=|=|=|=|=|}
% ~~~$\text{NiCl}_2$   &$-1.182$     &$0.51$  &$-0.176$  &$1.52$ &$0.07$ \cite{dagenais1976precise}  \\ \hline
% ~~~$\text{NiBr}_2$   &$-1.054$     &$0.48$  &$-0.176$  &$1.34$ &$0.31$ \cite{arnold1995study} \\ \hline
% ~~~$\text{NiI}_2$    &$-0.924$     &$0.6$ &$-0.215$  &$1.12$ &$0.63$ \cite{arnold1995study} \\ \hline
% \end{tabular}
% \end{centering}
% \caption{Hopping parameters and SOC terms in the $dp$ model for NiX$_{2}$ obtained from the maximally localized the 
% Wannier-basis sets in DFT. \textcolor{magenta}{How $\lambda_p$ was calculated ? The magnitude of the SOC $\lambda_p$ are taken from the atomic ${}^2P_{1/2}-{}^2P_{3/2}$ splitting measured by spectroscopy \cite{dagenais1976precise, arnold1995study}.} \textcolor{blue}{Note: These are written in slater-koster form and will be updated!}}
% \label{dp_model DFT parameters} 
% \end{table}
% \iffalse
\begin{table}[h]
\begin{centering}
\begin{tabular}{|l|l|l|l|l|l|}
\hline
 Parameters (eV) &~$t_{pd\sigma}~$ &~$t_{pp\sigma}$~  &~$t_{pp\pi}/t_{pp\sigma}$~  &~$\Delta_{pd}$~ &$\lambda_p$ \\ \hhline{|=|=|=|=|=|=|}
~~~$\text{NiCl}_2$   &$1.07$     &$0.48$  &$-0.18$  &$2.40$ &$0.07$ \cite{dagenais1976precise}  \\ \hline
~~~$\text{NiBr}_2$   &$0.95$     &$0.48$  &$-0.18$  &$1.92$ &$0.31$ \cite{arnold1995study} \\ \hline
~~~$\text{NiI}_2$    &$0.83$     &$0.55$ &$-0.21$  &$1.08$ &$0.63$ \cite{arnold1995study} \\ \hline
\end{tabular}
\end{centering}
\caption{Hopping parameters and SOC terms in the $dp$ model for NiX$_{2}$ obtained from the maximally localized the 
Wannier-basis sets in DFT. The ligand SOC $\lambda_p$ are taken from atomic spectroscopy results for
the ${}^2P_{1/2}$-${}^2P_{3/2}$ splitting \cite{dagenais1976precise, arnold1995study}.}
\label{dp_model DFT parameters} 
\end{table}

\section{Tight-binding Hamiltonian and dispersions for effective $e_g$-only model}\label{Appendix: Band Structure}

% \textcolor{red}{TSD: Shouldn't this section be after the present section III, which discusses construction of effective eg orbitals ?}

\subsection{Momentum space dispersion using effective tight-binding model parameters}
In this subsection, we provide the explicit momentum space Hamiltonian for the effective two-orbital $e_g$ model discussed in the main text.
As discussed above using DFT and using symmetry arguments in the main text,
the two $e_g$ orbitals in the effective model form a 2D representation of the approximate $O_h$ point group. Under a $2\pi/3$ rotation $C_{3z}$ about the crystallographic $z$-axis, they transform as
\begin{gather}
\begin{bmatrix}a^\dag \\ b^\dag \end{bmatrix} \to D_{\text{orb}}(C_{3z}) \begin{bmatrix}a^\dag \\ b^\dag \end{bmatrix} =   \begin{bmatrix}
-\frac{1}{2} &-\frac{\sqrt{3}}{2} \\ \frac{\sqrt{3}}{2} &-\frac{1}{2}
\end{bmatrix} \begin{bmatrix}a^\dag \\ b^\dag \end{bmatrix} .
\end{gather}
Including the spin rotation, the four-component spinor $\Psi^\dag = [a_{\ua}^\dag, a_{\da}^\dag, b_{\ua}^\dag, b_{\da}^\dag]$ transform as 
\begin{gather}
\Psi^\dag \to D_{\text{orb}}(C_{3z} ) \otimes \big( e^{-i\frac{\pi}{3} \sigma_z} \big) \Psi^\dag ,
\end{gather}
where $e^{-i\frac{\pi}{3}}$ is the $SU(2)$ representation of the $C_{3z}$ acting on spin. In momentum space, the Hamiltonian including both NN and 3NN hoppings are given by $H(\vec{k}) = H_{\text{hop}}(\vec{k}) + H_{\text{SOC}}(\vec{k}) =$. The hopping part from NN and 3NN is given by
\begin{equation}
    H_{\rm hop}(\vec{k}) =\begin{pmatrix} f_1(\bk) + f_3(\bk)-\mu  & h_1(\bk)+h_3(\bk) \\
   h_1(\bk)+h_3(\bk) & g_1(\bk) + g_3(\bk) - \mu
    \end{pmatrix} \otimes \sigma_0
\end{equation}
with
\begin{eqnarray*}
    f_1(\bk) &=& -2 t_{1a} \cos k_1 - \frac{1}{2} (t_{1a}+3 t_{1b}) (\cos k_2 + \cos k_3) \\
    f_3(\bk) &=& - 2 t_{3a} \cos 2 k_1 - \frac{1}{2} (t_{3a}+3 t_{3b}) (\cos 2 k_2 + \cos 2 k_3)\\
    g_1(\bk) &=& -2 t_{1b} \cos k_1 - \frac{1}{2} (3 t_{1a}+ t_{1b}) (\cos k_2 + \cos k_3) \\
    g_3(\bk) &=& - 2 t_{3b} \cos 2 k_1 - \frac{1}{2} (3 t_{3a}+ t_{3b}) (\cos 2 k_2 + \cos 2 k_3)\\
   h_1(\bk) &=& -\frac{\sqrt{3}}{2} (t_{1a}-t_{1b}) (\cos k_2 - \cos k_3) \\
    h_3(\bk) &=& -\frac{\sqrt{3}}{2} (t_{3a}-t_{3b}) (\cos 2 k_2 - \cos 2 k_3)
\end{eqnarray*}
Here, $k_1=k_x$, $k_2=-k_x/2+k_y \sqrt{3}/2$, and $k_3=-k_x/2-k_y \sqrt{3}/2$.  

The SOC-induced hopping part is given by
\begin{eqnarray}
    H^{\rm soc} &=& -2 i \lambda_{1z} \sum_\bk \left[(a^\dg_{\bk\uparrow} b^\pdg_{\bk\uparrow}
- a^\dg_{\bk\downarrow} b^\pdg_{\bk\downarrow}) (\cos k_1 + \cos k_2 + \cos k_3) \right] + h.c. \\
&+& 2 \lambda_{1y} \sum_\bk \left[(a^\dg_{\bk\uparrow} b^\pdg_{\bk\downarrow}
- b^\dg_{\bk\uparrow} a^\pdg_{\bk\downarrow}) (\cos k_1 + \omega \cos k_2 + \omega^2 \cos k_3) \right] + h.c.
%\\
%&-& 2 i \lambda_3 \sum_\bk \left[(a^\dg_{\bk\uparrow} b^\pdg_{\bk\uparrow}
%- a^\dg_{\bk\downarrow} b^\pdg_{\bk\downarrow}) (\cos 2 k_1 + \cos 2 k_2 + \cos 2 k_3) \right] + h.c.
\end{eqnarray}
where $\omega=e^{i 2\pi/3}$. We plot below representative dispersions without and with SOC hopping terms,
showing band touching points for the case of zero SOC which get gapped out by turning on SOC terms.
% \end{widetext}

\begin{figure}[h]
    \centering
    \includegraphics[width=0.8\linewidth]{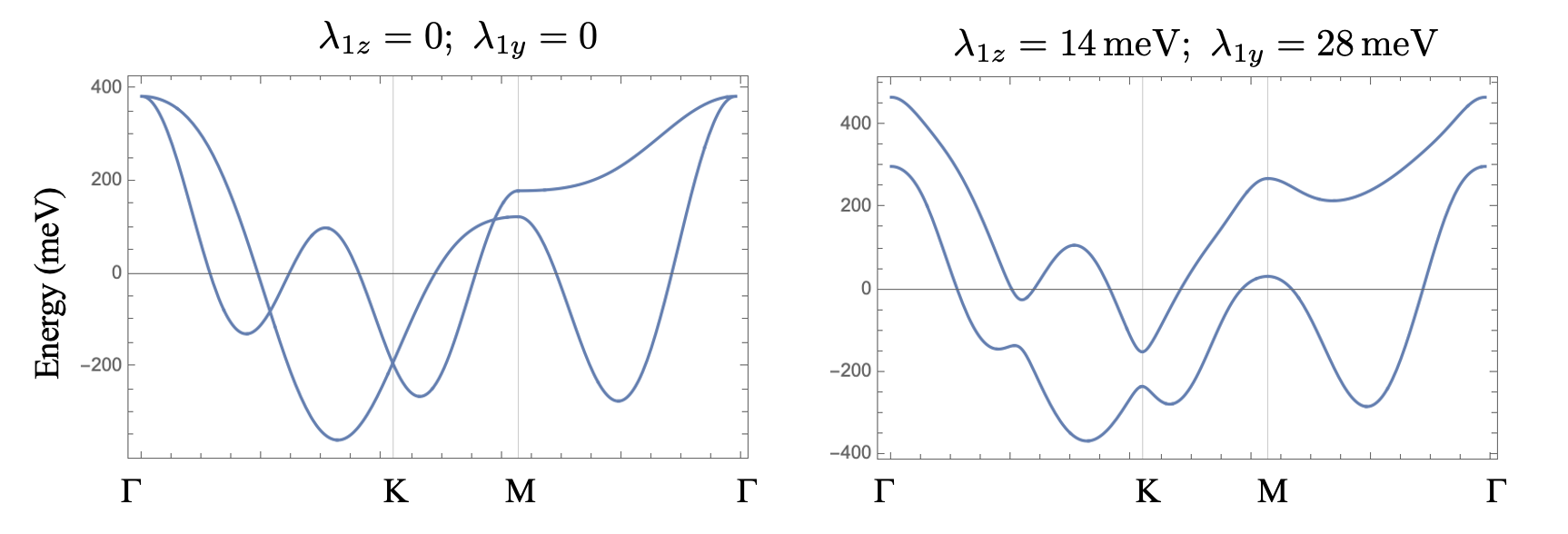}
    \caption{Two-orbital tight-binding dispersions for parameters $t_{1a}=-22$\,meV, $t_{1b}=-36$\,meV,
    $t_{3a}=25$\,meV, $t_{3b}=-95$\,meV from Table I of main text for {NiI$_2$}. Each band is two-fold
    degenerate due to inversion and time-reversal symmetry. Left panel shows dispersion
    in the absence of SOC hopping terms which exhibits band touching points; these degeneracies are
    lifted by SOC terms (right panel) where we have chosen $\lambda_{1z}$, $\lambda_{1y}$ from Table I of main text for {NiI$_2$}.}
    \label{fig:bandtouchingSM}
\end{figure}

\subsection{Microscopic insights into the hopping parameters}
In the main text, we have provided qualitative insights into the hierarchy of energy scales associated with
first, second, and third neighbor hoppings as well
as ligand induced SOC. 
In this subsection, we provide a partial microscopic understanding of the hopping parameters in our effective two-orbital model, i.e, $(t_{1a}, t_{1b}, t_{3a}, t_{3b}, \lambda_{1z}, \lambda_{1y})$. Based on the DFT parameters in Table I of main text, we infer that the direct hopping between $d$-orbitals is negligible in the family of $\text{NiX}_2$, since $(t_{1a}, t_{1b}, t_{3a}, t_{3b})$ are smallest for $\text{NiCl}_2$ and increase in $\text{NiBr}_2$ and $\text{NiI}_2$. The increase of the hopping parameters from $\text{Cl}\to \text{Br} \to \text{I}$ is consistent with enhanced $p\text{-}d$ hybridization, while the direct $d-d$ hoppings are expected to decrease from $\text{Cl}\to \text{Br} \to \text{I}$ because of the increasing bond length.

For simplicity, throughout the Supplementary Material, we focus on ideal octahedra and ignore the effects of distortion. We also assume that orbitals on different atomic sites are orthogonal. We expect distortions to play only a minor role in $\mathrm{NiX_2}$ materials. However, the second assumption is stronger, since ligand $p$ orbitals are spatially extended and can have non-negligible overlap with neighboring ligand and metal orbitals.

To obtain the ligand-$p$-mediated hoppings, we start from a $dp$-Hamiltonian that includes nearest-neighbor Ni-X hoppings, and nearest-neighbor X-X hoppings in term of Slater Koster parameters $(t_{pd\sigma}, t_{pp\sigma}, t_{pp\pi})$ and onsite energies $(\lambda_p, \varepsilon_p, \varepsilon_{e_g})$. Integrating out the ligand $p$ orbitals yields an energy-dependent effective Hamiltonian in the $d$-subspace. We evaluate this Hamiltonian at the energy of the antibonding $e_g^*$ state $E = -E_{e_g^*}$ (formed from strong Ni-$X$
hybridization), and define $\Delta_{pd} = \varepsilon_{e_g} - \varepsilon_p$,
\begin{gather}
\Delta_{*} = E_{e_g^*} - \varepsilon_p = \frac{\big(\Delta_{pd} + (t_{pp\sigma}-t_{pp\pi}) \big)}{2} + \sqrt{\left( \frac{\big(\Delta_{pd} - (t_{pp\sigma}-t_{pp\pi}) \big)}{2} \right)^2 + 3t_{pd\sigma}^2} .
\end{gather} 
In the limit where the ligand SOC $\lambda_p \ll \Delta_{*}$, we summarize the simplified expressions for the
leading order perturbative results, including M-X-M and M-X-X-M paths, for the spin-independent and spin-dependent hoppings.

(i) For the NN hoppings along $\vec{\delta}_1$-bond,
 \begin{gather}
t_{1a} \approx -\frac{t_{pd\sigma}^2(t_{pp\sigma}+9t_{pp\pi})}{4\Delta_{*}^2},~~\quad t_{1b} \approx -\frac{3t_{pd\sigma}^2(t_{pp\sigma}+t_{pp\pi})}{4\Delta_{*}^2}, \\
\lambda_{1z} \approx \lambda_p \left(\frac{t_{pd\sigma}^2}{4\Delta_{*}^2} + \frac{2t_{pd\sigma}^2(t_{pp\sigma} - t_{pp\pi})}{4\Delta_{*}^3} \right),~~
\lambda_{1y} = \sqrt{2} \lambda_{1z} .
 \end{gather} 

(ii) For the 2NN hoppings along the $(\delta_2+\delta_3)$-bond, the intralayer X-X paths add destructively, while only one interlayer X-X path contributes to the spin-independent hoppings, yielding
\begin{gather}
t_{2a} = \frac{t_{pd\sigma}^2 t_{pp\pi}}{\Delta_*^2}, \quad t_{2b} = 0, \\
\lambda_{2z} \approx \lambda_p \frac{t_{pd\sigma}^2(t_{pp\sigma}-5t_{pp\pi})}{4\sqrt{2}\Delta_*^3}, \quad \lambda_{2y} \approx \lambda_p \frac{t_{pd\sigma}^2(t_{pp\sigma}+t_{pp\pi})}{2\Delta_*^3} .
\end{gather}

(iii) For the 3NN hoppings along the $\vec{\delta}_1$-bond,
 \begin{gather}
t_{3b} \approx -\frac{3t_{pd\sigma}^2(t_{pp\sigma}-t_{pp\pi})}{4\Delta_{*}^2}, \quad t_{3a} = -\frac{1}{3}t_{3b}, \\
\lambda_{3z} \approx \lambda_p\frac{t_{pd\sigma}^2(2 t_{pp\sigma}+ t_{pp\pi})}{2\Delta_{*}^3}, \quad 
\lambda_{3y} = \sqrt{2}\lambda_{3z},
 \end{gather}
Thus, spin-independent 2NN hoppings are suppressed compared to 1NN, 3NN terms by factor $\sim t_{pp\pi}/t_{pp\sigma} \ll 1$. Similarly, SOC hoppings for 2NN and 3NN, are smaller that 1NN SOC terms ($\lambda_{1y}$, $\lambda_{1z}$) 
by factor $\sim t_{pp\sigma}/\Delta_* \ll 1$.
Spin-{\it independent} hoppings are mediated by $X$-$X$ hoppings which require finite $t_{pp\sigma}$ or $t_{pp\pi}$. However, 
1NN spin-{\it dependent} hoppings do not require finite $t_{pp\sigma}$ or 
$t_{pp\pi}$, although their magnitudes are enhanced by $X$-$X$ hoppings.

Using the above perturbation theory (PT) equations, and the DFT derived $dp$ parameters listed in Table.~\ref{dp_model DFT parameters}, we estimate the effective hopping parameters for the $e_g$ model. 
These results are summarized as `PT' in Table.~\ref{Perturbative estimation of parameters}. 
% We find that the hierarchy of hoppings with larger $t_{1b},t_{3b}$, and presence of 
% nonzero SOC hoppings $\lambda_{1z},\lambda_{1y}$ are in qualitative agreement with the DFT results in Table I of the main text.
% We find that the 1NN hopping parameters $t_{1a}, t_{1b}, \lambda_{1z}, \lambda_{1y}$ are slightly overestimated compared to the DFT-estimated values, whereas the 3NN hoppings show good quantitative agreement.
% We attributed this discrepancy for the 1NN parameters
% to the extended nature of the DFT Wannier orbitals, so these parameters are not fully captured within the PT treatment.

\begin{table}[h]
\begin{centering}
\begin{tabular}{|l|l|l|l|l|l|l|l|l|l|}
\hline
(meV) &~$t_{1a}~$ &~$t_{1b}$~  &~$t_{3a}$~  &~$t_{3b}$~  &~$\lambda_{1z}$~ &~$\lambda_{1y}$  &~$\lambda_{3z}$~ &~$\lambda_{3y}$~\\ \hhline{|=|=|=|=|=|=|=|=|=|}
~~~$\text{NiCl}_2$(PT) &$7$ &$-27$ &$7$ &$-39$ &$2$ &$3$ &$0.8$ &$1.1$   \\ \hline
~~~$\text{NiCl}_2$(DF) &$5$ &$-39$ &$17$ &$-52$ &$2.4$ &$3.3$ &$0.3$ &$0.3$   \\ \hline
~~~$\text{NiBr}_2$(PT)  &$7$     &$-29$  &$14$  &$-42$  &$10.5$ &$14.9$ &$4.4$ &$6.2$   \\ \hline
~~~$\text{NiBr}_2$(DF)  &$4$     &$-48$  &$20$  &$-61$  &$12$ &$17$ &$2$ &$2$   \\ \hline
~~~$\text{NiI}_2$(PT)    &$15.6$     &$-42$  &$22$  &$-63$  &$31.5$ &$44.6$ &$17$ &$24$ \\ \hline
~~~$\text{NiI}_2$(DF)    &$3$     &$-113$  &$41$  &$-137$  &$53$ &$63$ &$13$ &$11$ \\ \hline
\end{tabular}
\end{centering}
\caption{Perturbative treatment (PT) estimated hopping parameters in comparison to DFT downfolding (DF) estimates.}
\label{Perturbative estimation of parameters}
\end{table}

\begin{figure}[h]
\centering
\includegraphics[width=0.85\textwidth]{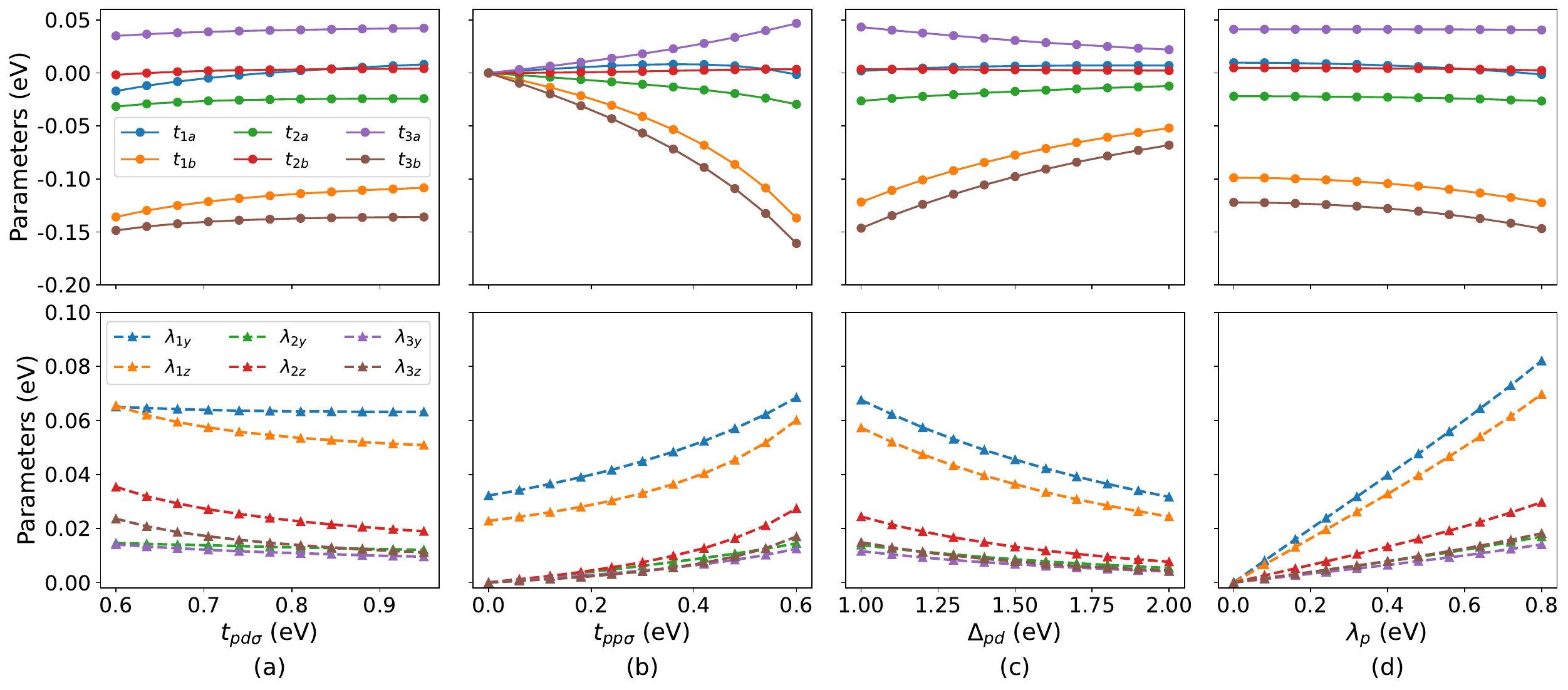}
\caption{Variation of exact downfolded $e_g$ hopping parameters obtained as functions of (a) $t_{pd\sigma}$, (b) $t_{pp\sigma}$, (c) $\Delta_{pd}$ and (d) $\lambda_p$. The spin-independent and spin-dependent (SOC induced) hopping parameters are shown respectively in the top panels (solid line) and bottom panels (dash lines). In panels (a), (b), (c) and (d) we vary the relevant parameter while respectively fixing the other parameters
to [ $t_{pp\sigma} = 0.55$\,eV, $t_{pp\pi}/t_{pp\sigma} = -0.21$, $\Delta_{pd} = 1.08$\,eV, $\lambda_p = 0.63$\,eV ], 
[ $t_{pd\sigma} = 0.83$\,eV, $t_{pp\pi}/t_{pp\sigma} = -0.21$, $\Delta_{pd} = 1.08$\,eV, $\lambda_p = 0.63$\,eV ], 
[$t_{pd\sigma} = 0.83$\,eV, $t_{pp\pi}/t_{pp\sigma} = -0.21$, $\lambda_p = 0.63$\,eV ] 
and [ $t_{pd\sigma} = 0.83$\,eV, $t_{pp\sigma} = 0.55$\,eV, $t_{pp\pi}/t_{pp\sigma} = -0.21$, $\Delta_{pd} = 1.08$\,eV], 
respectively.}
\label{Exact downfolding parameters eg}
\end{figure}

To incorporate the role of higher-order ligand-mediated processes, we further examine the hopping amplitudes using an exact downfolding 
(DF) procedure starting from the $dp$ model, which incorporate X-X processes to all orders. The resulting `DF' values for $\mathrm{NiX_2}$
using DFT parameters for the $dp$ model are also listed in Table.~\ref{Perturbative estimation of parameters}.  
Finally, present plots of the exact downfolding results for the $dp$ model to obtain the dependence of the effective $e_g$ model
parameters on $t_{pd\sigma}, t_{pp\sigma}, \Delta_{pd}, \lambda_p$ as shown in 
Fig.~\ref{Exact downfolding parameters eg}, This provides a convenient framework for analyzing the evolution of effective 
hoppings under external tuning parameters such as pressure or strained in the {1T-M$X_2$} materials. 
We note that these analytical perturbative (PT) and downfolding (DF) results for the effective 
two-orbital model parameters starting from an atomic $dp$ model roughly capture the trends in Table I of the main text
in terms of showing why $t_{3b}, t_{1b}$ are dominant hopping parameters, and correctly describing 
the relative strength of the SOC hopping terms
$\lambda_{1z},\lambda_{1y}$.

\section{Topology, Edge/Corner states}

In this section, we include additional results related to the first order topological bands (spin Chern bands) and higher order topological bands discussed in the main text.

\subsection{Spin Chern bands and Corner charges for $C^s_{-}=+2$}
\begin{figure}[H]
\centering
\subfloat[]{\includegraphics[width=0.48\textwidth]{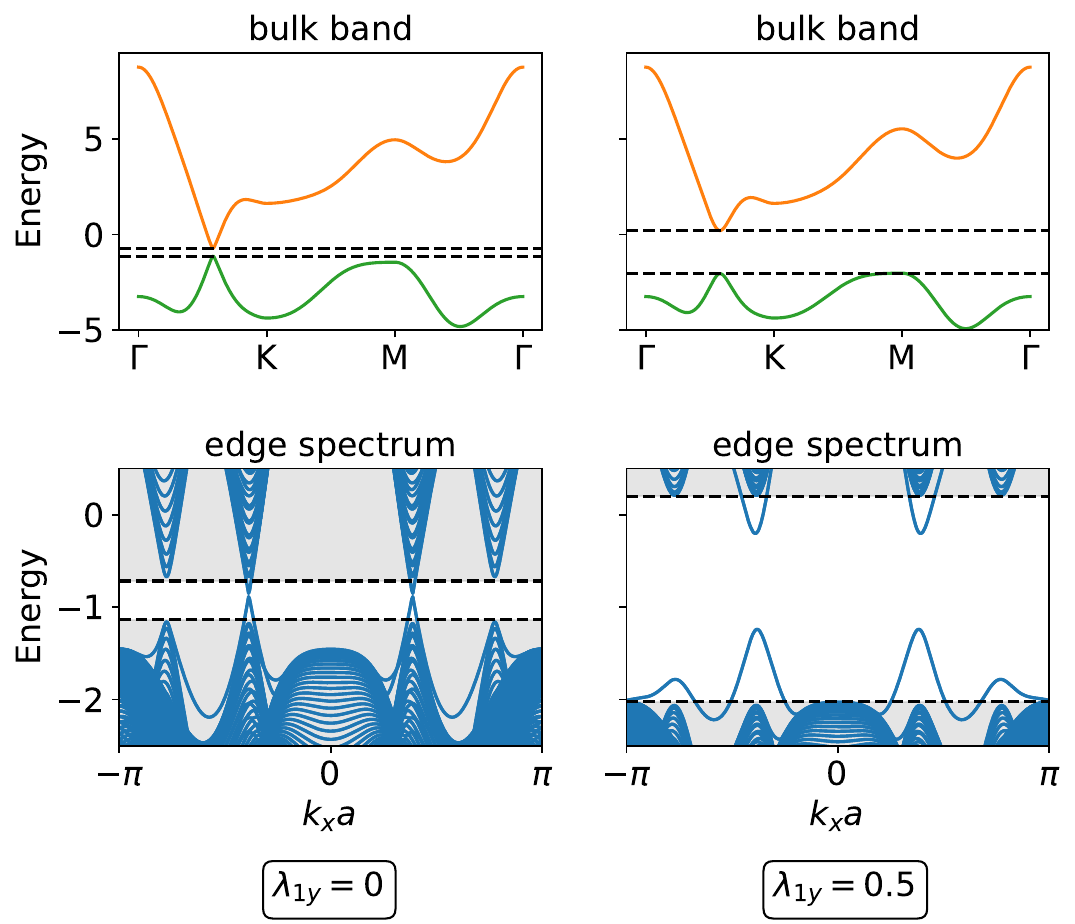}\label{SMspinchern-c2}}
\subfloat[]{\includegraphics[width=0.48\textwidth]{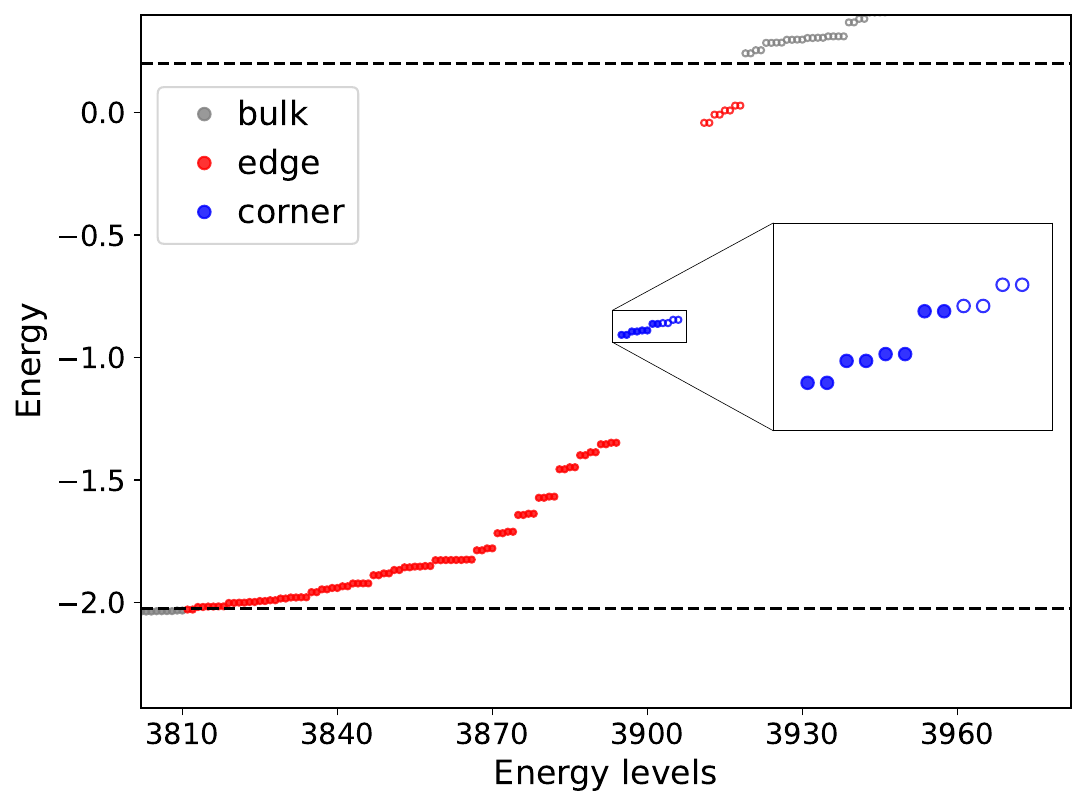}\label{SMHOT-c2}}
\caption{(a) Gapped bulk band dispersion and edge spectrum for $C^s_{-} = +2$ for spin-mixing SOC $\lambda_{1y}=0$ (left) and
$\lambda_{1y}=0.5$ (right). A nonzero $\lambda_{1y}$ enhances the bulk band gap of the topological insulator,
and gaps out the edge states. (b) Gapped bulk band dispersion in the HOT phase starting from $C^s_{-}=+2$. We see clear gap between bulk
valence and conduction bands, edge modes (red), and corner localized modes (blue). For the corner modes,
we find 8 out of 12 states being occupied at half-filling, which would correspond to a fractional corner charge
on a hexagonal flake which is
$\mathcal{Q}_c = 4e/3 \text{ mod } 2e$ per corner.}
\end{figure}

In the main text, we have shown the bulk bands and edge modes for the case $C^{s}_{-} = -4$. In Fig.~\ref{SMspinchern-c2} and Fig.~\ref{SMHOT-c2}, we show similar plots of edge and corner states, respectively, for the case of $C^{s}_{-} = +2$.

% \subsection{Fractional corner anomaly for higher order topological bands}

% In the main text, we have discussed the example of cases with nontrivial higher order topological indices where the corner mode energies
% overlap with the bulk valence band energies. In  such that it is not possible to spectrally isolate them or study their
% corner localization. In this case, we resorted to the fractional corner anomaly as a way to look for these corner modes. 
% \textcolor{blue}{Some details and figs related to how FCA is extracted.}

% Here, we
% show the eigenvalue spectrum for this case, which we refer to in the main text.
% We consider the parameter set $(t_{1a}, t_{1b}, t_{3a}, t_{3b}, \lambda_{1z}, \lambda_{1y}) = (0, 0.3, 1/3, -1, 1, 0.5)$ for the calculations. Diagonalizing on a finite hexagonal flake with $N_{\text{site}} = 1951$, we obtain two isolated bands, as shown in Fig.~\ref{FCA example}.
% The marked red edge modes merge with the upper $+$ band. The corner modes sink into the bulk valence band and are not detectable.

% \begin{figure}[H]
% \centering
% \includegraphics[width=0.48\textwidth]{Figures/Topology_Figures/Corner_states_inband.pdf}
% \caption{Energy spectrum of a finite hexagonal flake with $N_{\text{site}} = 1951$. Bulk states and edge states are distinguished by green and purple, respectively. The parameters are $(t_{1a}, t_{1b}, t_{3a}, t_{3b}, \lambda_{1z}, \lambda_{1y}) = (0, 0.3, 0, -1, 1, 0.5)$. The inset shows the density of states (DOS), which indicates two separated bands labeled by (-) and (+).}
% \label{FCA example}
% \end{figure}

\subsection{Topological index and Fractional corner anomaly}
In the main text, we focus on a parameter regime in which the bulk spectrum is already gapped at $\lambda_{1y} = 0$, with gapless edge Dirac cone appearing within the bulk gap. Turning on $\lambda_{1y}$ gaps out the edge Dirac cones, leading to the emergence of corner states within the resulting edge gap \cite{hung2024time}.
By contrast, as pointed out in the main text,
there also exist regimes in which the bulk remains gapless at $\lambda_{1y}=0$. In such cases, the edge Dirac cones appear within a bulk band, and the resulting corner states are embedded in the bulk continuum. Even if a bulk gap is subsequently opened upon increasing $\lambda_{1y}$, corner states are not guaranteed to lie within the bulk gap unless additional protecting symmetries are present.

\begin{figure}[H]
\centering
\subfloat[Energy spectrum]{\includegraphics[width=0.48\textwidth]{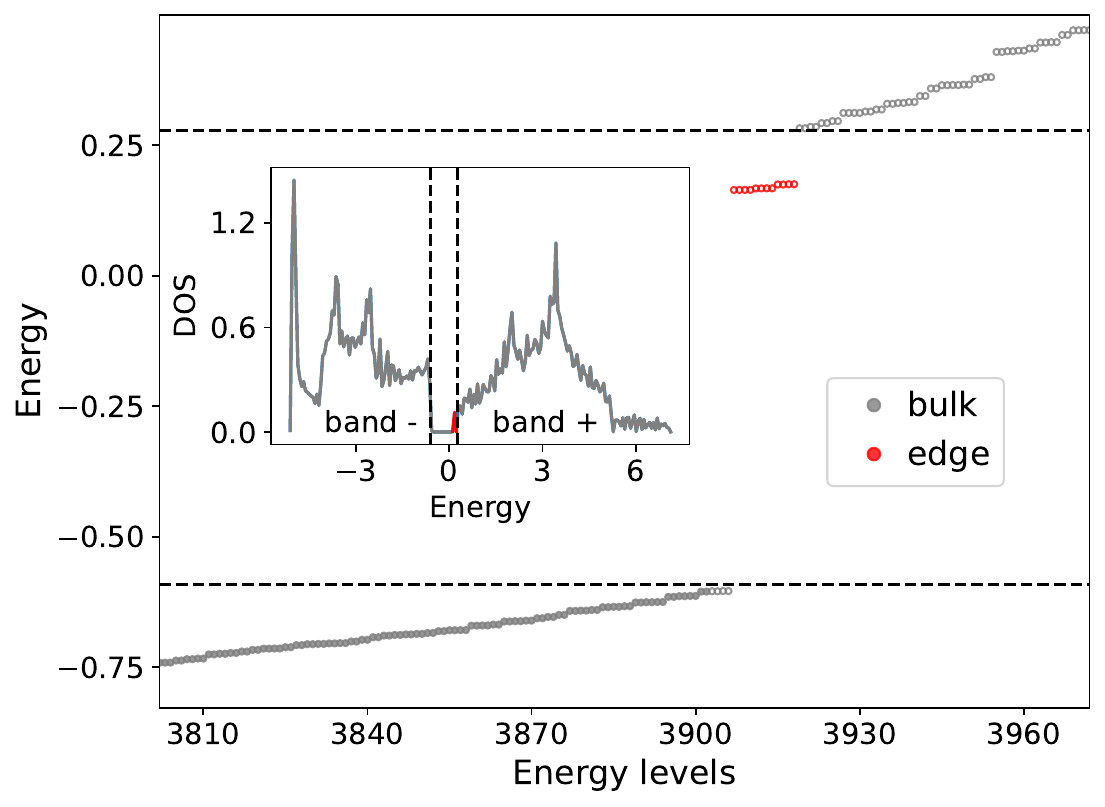}\label{FCA_example}}
\subfloat[Plot of the lower band $\Phi_{-}(i)$]{\includegraphics[width=0.4\textwidth]{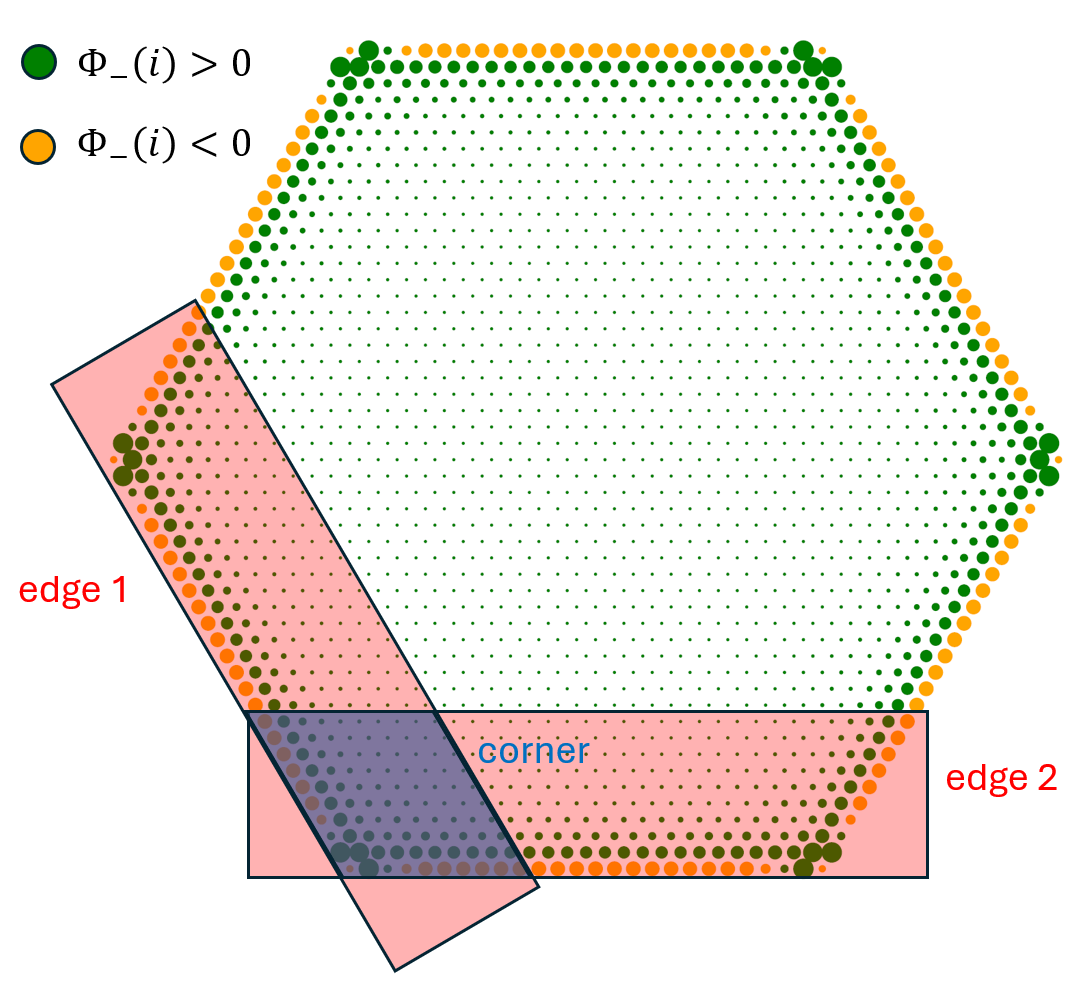}\label{FCA_schematic}}
\caption{(a) Energy spectrum of a finite hexagonal flake with $N_{\text{site}} = 1951$. Bulk states and edge states are distinguished by gray and red, respectively. The parameters are $(t_{1a}, t_{1b}, t_{3a}, t_{3b}, \lambda_{1z}, \lambda_{1y}) = (0, 0.3, 0, -1, 1, 0.5)$. The inset shows the density of state(DOS), which indicate two separated bands labeled by (-) and (+). (b) Spatial distribution of the  $\Phi_{band}(i)$ for {\it band} set to lower (-) band. Green and orange dots denote positive and negative values of $\Phi_{band}(i)$, respectively, where the size of the dots indicate the magnitude. The purple $N_L\times N_L = 10\times 10$ parallelogram denotes the corner region used to compute $\phi_-$, while the two rectangular boxes indicate two adjacent edges which share the same corner.}
\end{figure}

To illustrate its relation to HOT and its connection with the fractional corner charge $\mathcal{Q}_c$, we consider the parameter set $(t_{1a}, t_{1b}, t_{3a}, t_{3b}, \lambda_{1z}, \lambda_{1y}) = (0, 0.3, 1/3, -1, 1, 0.5)$ for the calculations. Diagonalizing on a finite hexagonal flake with $N_{\text{site}} = 1951$, we obtain two isolated bands, as shown in 
Fig.~\ref{FCA_example}, with
mid-gap edge modes but no mid-gap corner modes. We thus
follow Ref.~\cite{peterson2020fractional} to calculate the fractional corner anomaly (FCA), for which results are presented in the
main text.

To compute the fractional corner anomaly (FCA) in the main text, we first define the band-resolved local charge density with the uniform background charge subtracted,
\begin{gather}
\Phi_{\text{band}}(i) \equiv \sum_{n\in \text{band}} |\psi_{n}(i)|^2 - \frac{N_{\text{band}}}{N_{\text{site}}},
\end{gather}
where $i$ labels lattice sites, $N_{\text{band}}$ is the number of states in the selected band, $N_{\text{site}}$ is the number of sites, and $\psi_n(i)$ denotes the wavefunction of the $n$-th eigenstate on site $i$. This subtraction removes the spatially uniform contribution, such that $\sum_i \Phi_{\text{band}}(i) = 0$. Setting $band$ to the lower (-) band in $\Phi_{band}(i)$, we compute the FCA \cite{peterson2020fractional} as
\begin{gather}
\phi_{-} = \rho_{\text{corner}} - (\sigma_1 + \sigma_2) \quad \text{ mod 2e}.
\end{gather}
Here $\rho_{\text{corner}}$ is obtained by summing $\Phi_{band = (-)}(i)$ over the purple $N_L\times N_L = 10\times 10$ parallelogram in Fig.~\ref{FCA_schematic}. The quantities $\sigma_1$ and $\sigma_2$ denote edge contributions associated with the two adjacent edges of the same corner. More explicitly, the two first-order topological edges intersect at the purple region and contribute corner-localized density equal to $\sigma_1+\sigma_2$ in this region, and the deviation $\phi_-$ from this value indicates nontrivial HOT. For each adjacent edge, the corresponding edge contribution is estimated layer by layer. For a given layer $j$, we take the value $\Phi_{band=(-)}(i)$ at the center of the edge as the representative charge density for the entire layer. The contribution from this layer to the $N_L\times N_L = 10\times 10$ corner region is then given by
\begin{gather}
\sigma_1 = N_L \sum_{j=1}^{N_L} \Phi^{(1)}_{\text{center}}(j), \quad \sigma_2 = N_L \sum_{j=1}^{N_L} \Phi^{(2)}_{\text{center}}(j),
\end{gather}
where $\Phi^{(1)}_{\text{center}}(j)$ and $\Phi^{(2)}_{\text{center}}(j)$
denote the $j\text{th}$-layer of edge 1 and 2, respectively.

\subsection{Corner states in diamond-shaped nanoflake}
\begin{figure}[h]
\centering
\includegraphics[width=0.48\textwidth]{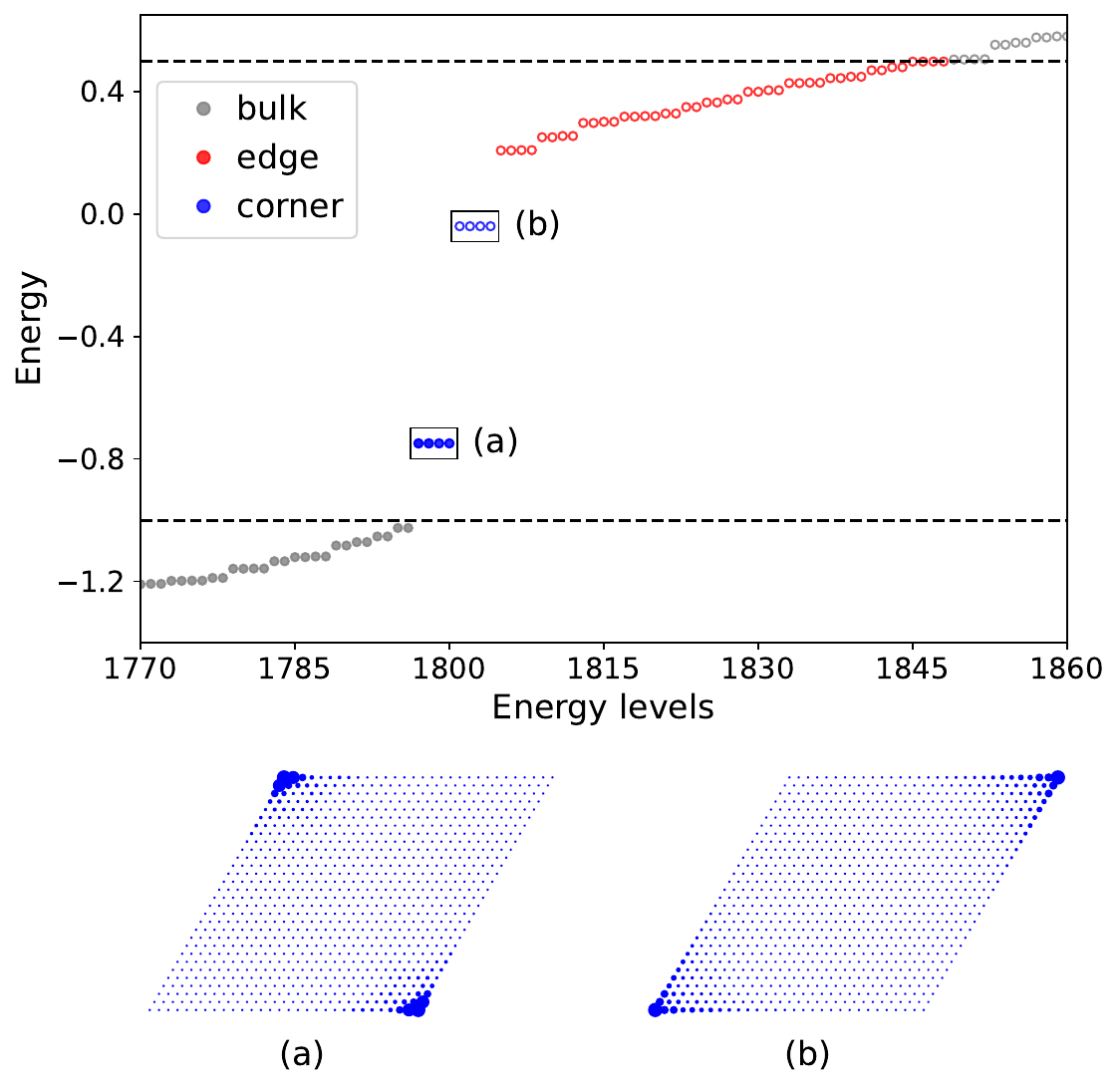}
\caption{Energy spectrum of a $45\times 45$ diamond-shaped nanoflake using the same $C_-^s = -4$ parameters labeled in Fig. 4. Panels (a) and (b) show the local charge density of the states marked by the corresponding blue boxes in the spectrum.}
\label{corner states in diamond-shaped nanoflake}
\end{figure}

The corner states discussed in the main text are not limited to a specific nanoflake geometry. In Fig.~\ref{corner states in diamond-shaped nanoflake}, we show the energy spectrum for the $C_-^s = -4$ example using a $45\times 45$ diamond-shaped nanoflake. The spectrum exhibits in-gap corner-localized states, demonstrating that the corner modes persist in a different nanoflake shape. The resulting $\phi_-$ is changed by $< 0.01$ by varying from $N_L =8$ to $N_L = 12$.

\section{Spin Model from the two-orbital Hubbard-Kanamori model}\label{Appendix: Spin Model derivations}
We consider the local Hubbard-Kanamori Hamiltonian at site $i$ for $e_g$ orbitals, which is given by
\begin{gather}
\begin{aligned}
H_{\text{int}} &=U \sum_{\ell} n_{i\ell\ua} n_{i\ell\da} +  U'\sum_{\ell \neq  \ell'} n_{i \ell \ua} n_{i\ell \da} + (U'-J_H)\sum_{\ell<\ell',\sigma} n_{i\ell\sigma} n_{i\ell'\sigma}
+ J_H \sum_{\ell \neq \ell'} c_{i\ell\ua}^{\dag} c_{i\ell'\da}^\dag c_{i\ell\da} c_{i\ell'\ua} + J' \sum_{\ell \neq \ell'} c_{i\ell \ua}^\dag c_{i\ell \da}^\dag c_{i\ell'\da} c_{i\ell'\ua}\\
   &= U \sum_{\ell} n^\pdg_{i\ell \upa}  n^\pdg_{i\ell\dna} + V \sum_{\ell < \ell'} n^\pdg_{i\ell} 
    n^\pdg_{i\ell'} 
 -    J_H \sum_{\ell \neq \ell'} \bS^\pdg_{i\ell} \cdot \bS^\pdg_{i\ell'}
   + J' \sum_{\ell\neq \ell'} c^\dg_{i\ell\upa}
    c^\dg_{i\ell\dna} c^\pdg_{i\ell'\dna} c^\pdg_{i\ell'\upa}
\end{aligned}
\end{gather}
Here $U, U'$ denote the intra- and interorbital interactions, respectively. $J_H$ is the Hund's coupling, and $J'$ is the pair-hopping amplitude. For simplicity, we denote $V = (U'-J_H/2)$.
We list in Table.~\ref{single site energy for Hubbard-Kanamori Hamiltonian} the single-site eigenstates of the local Hamiltonian, grouped by electron filling $n$, and their energies.

\setlength{\tabcolsep}{6pt}
\renewcommand{\arraystretch}{1.15}

\begin{table}[H]
\begin{ruledtabular}
\begin{tabular}{c l c}
$n$ & $\lvert\psi\rangle$ & $E$ \\ \hline

1 &
$\begin{aligned}
&\lvert\uparrow,0\rangle,\ \lvert\downarrow,0\rangle,\\
&\lvert0,\uparrow\rangle,\ \lvert0,\downarrow\rangle
\end{aligned}$
& $0$ \\
\hline
2 & Triplet ($S=1$):\ 
$\begin{aligned}
&\lvert\uparrow,\uparrow\rangle,\ \lvert\downarrow,\downarrow\rangle,\\
&\tfrac{1}{\sqrt2}\big(\lvert\uparrow,\downarrow\rangle+\lvert\downarrow,\uparrow\rangle\big)
\end{aligned}$
& $V - \frac{J_H}{2} =U'-J_H$ \vspace{1em} \\

  & Singlet ($S=0$):\
$\tfrac{1}{\sqrt2}\big(\lvert\uparrow,\downarrow\rangle-\lvert\downarrow,\uparrow\rangle\big)$
& $V + \frac{3J_H}{2} =  U'+J_H$ \vspace{1em} \\

  & Doubly occupied:
$\begin{aligned}
&\tfrac{1}{\sqrt2}\big(\lvert\uparrow\downarrow,0\rangle-\lvert0,\uparrow\downarrow\rangle\big)\\
&\tfrac{1}{\sqrt2}\big(\lvert\uparrow\downarrow,0\rangle+\lvert0,\uparrow\downarrow\rangle\big)
\end{aligned}$
& $\begin{aligned}U-J' \\ U+J' \end{aligned}$ \\
\hline
3 &
$\begin{aligned}
&\lvert\uparrow,\uparrow\downarrow\rangle,\ \lvert\downarrow,\uparrow\downarrow\rangle,\\
&\lvert\uparrow\downarrow,\uparrow\rangle,\ \lvert\uparrow\downarrow,\downarrow\rangle
\end{aligned}$
& $U+2V = U + 2 U' - J_H$ \\
\hline
4 & $\lvert\uparrow\downarrow,\uparrow\downarrow\rangle$ & $2 U + 4 U'$ \\

\end{tabular}
\end{ruledtabular}
\caption{Single-site eigenstates and energies of the $e_g$ orbitals Hubbard--Kanamori interaction. 
We use $\lvert s_1,s_2\rangle$ to denote a state with spin $s_1$ in orbital $a$ and spin $s_2$ in orbital $b$.}
\label{single site energy for Hubbard-Kanamori Hamiltonian}
\end{table}
At half-filling of the two-orbital model, the low-energy subspace is formed by the triplet states with $S = 1$. In order to derive an effective spin model, we use our effective tight-binding model in Eq.~\ref{NN electron hopping matrix along x} for NN, and \ref{3NN electron hopping along x} for 3NN. We assume a strong coupling limit where $U,J_H \gg t_{1a}, t_{1b}, t_{3a}, t_{3b}, \lambda_{1z}, \lambda_{1y}$, and obtain the effective Hamiltonian in the low-energy subspace using second-order perturbation theory. The resulting exchange matrix for $\vec{\delta}_1$-bond is
\begin{gather}
\label{Exchange parameters in global coordinate}
\begin{aligned}
\mathbf{J}_{1}^{\vec{\delta}_1-{\mathrm{bond}}} &= \frac{1}{U+J_H} \times  
\begin{bmatrix} t_1^2 - 2\lambda_{1z}^2 -2\lambda_{1y}^2 &0 &0 \\ 0 &t_1^2-2\lambda_{1z}^2+2\lambda_{1y}^2 &-4\lambda_{1z} \lambda_{1y} \\ 0 &-4\lambda_{1z} \lambda_{1y} &t_1^2+2\lambda_{1z}^2 - 2\lambda_{1y}^2
\end{bmatrix} \\
\mathbf{J}_3 &= \frac{t_3^2}{U+J_H} \mathbb{I}
\end{aligned}
\end{gather}
where $t_1^2\equiv t_{1a}^2 + t_{1b}^2$ and $t_3^2\equiv t_{3a}^2 + t_{3b}^2$. The other bonds are related by a $C_3$ rotation along the crystallographic $z$-axis. Rotating the spin basis to the local octahedral coordinate, the exchange Hamiltonian can be rewritten as the extended Kitaev-Heisenberg $\mathrm{J-K-\Gamma-\Gamma'}$ with conversion explained
in Eq.~\eqref{Exchange parameters conversion}. The two parameterizations are related by
\begin{gather}
\label{Exchange parameters conversion}
\begin{aligned}
J&= \frac{1}{6}(3J_{xx} + J_{yy} + 2J_{zz} + 2\sqrt{2}J_{yz}) \\
K&= \frac{1}{2}(-J_{xx}+ J_{yy} - 2\sqrt{2}J_{yz}) \\
\Gamma &= \frac{1}{6}(-3J_{xx} + J_{yy} + 2J_{zz} + 2\sqrt{2}J_{yz}) \\
\Gamma' &= \frac{1}{6}(-2J_{yy} + 2J_{zz} - \sqrt{2}J_{yz}) .
\end{aligned}
\end{gather}
Using this, we obtain the exchange parameters in terms of $J_1,K_1,\Gamma_1,\Gamma'_1$ as given in the main text.

\section{Magnetoelectricity}

In this section, we provide microscopic insights and derivation of the magneto-electric effect discussed in the main text,
which extends the generalized Katsura-Nagaosa-Balatsky framework to the two-orbital $e_g$ case. This needs four main steps: (A) the
construction of Ni$X_6$ molecular orbital wavefunctions as an approximation to the Wannier wavefunction;
(B) the inclusion of SOC on the ligand $X$ which modifies these
molecular orbital wavefunctions leading to spin-orbit tails; (C) writing the effective polarization operators on Ni-Ni
bonds in terms of an effective electron position operator; and finally (D) using these modified wavefunctions to 
calculate the effective position operator and thus the effective polarization on the Ni-Ni bonds.
% \section{Microscopic insights into the effective $e_g$ model tight-binding parameters}\label{Appendix: Microscopic discussions}

\subsection{Construction of the effective molecular $e_g$ orbitals}
The effective two-orbital model is formed from strongly hybridized ligand states with $e_g$ symmetry, rather than the atomic Ni $e_g$ orbitals. To illustrate this point, we focus on the local $\mathrm{NiX_6}$ octahedron, which capture the dominant $p\mathrm{-}d$ hybridization around a Ni site. Within this cluster, only symmetry-adapted linear combinations of $p$-orbitals couples to the central Ni $e_g$ and $t_{2g}$. We denote these ligand states by $|P^{E_g}_{\alpha}\rangle$ and $|P^{T_{2g}}_{\alpha}\rangle$, respectively. \\

For convenience, we use the basis of (site, orbital, spin) indices, where the ligand-site index is ordered as $(+\hat{X}, -\hat{X}, +\hat{Y}, -\hat{Y}, +\hat{Z}, -\hat{Z})$. We define
\begin{gather}
|X_-\rangle = (1, -1, 0,0,0,0), \quad |Y_-\rangle = (0,0,1,-1,0,0), \quad |Z_-\rangle = (0,0,0,0, 1, -1).
\end{gather}
The symmetry-adapted ligand states are then given by
\begin{gather}
\begin{aligned}
|P_{X^2-Y^2}^{E_g}, \sigma\rangle &= \frac{1}{2} \big[ |X_-\rangle |p_{X},\sigma \rangle - |Y_-\rangle |p_{Y},\sigma \rangle \big], \\
|P_{3Z^2-R^2}^{E_g}, \sigma \rangle &= \frac{1}{\sqrt{12}} \big[ -|X_-\rangle |p_{X}, \sigma \rangle - |Y_-\rangle |p_{Y}, \sigma \rangle + 2|Z_-\rangle |p_{Z}, \sigma \rangle \big] , \\
|P_{XY}^{T_{2g}}, \sigma \rangle &= \frac{1}{2} \big[ |X_-\rangle |p_{Y}, \sigma \rangle + |Y_-\rangle |p_{X}, \sigma\rangle \big], \\
|P_{YZ}^{T_{2g}}, \sigma \rangle &= \frac{1}{2} \big[|Y_-\rangle |p_{Z}, \sigma \rangle + |Z_-\rangle |p_{Y}, \sigma \rangle \big], \\
|P_{XZ}^{T_{2g}}, \sigma \rangle &= \frac{1}{2} \big[ |X_-\rangle |p_{Z}, \sigma\rangle + |Z_-\rangle |p_{X}, \sigma\rangle \big].
\end{aligned}
\end{gather}
where $\sigma$ denotes spin. \\

Their corresponding ligand energies are $E(P^{E_g}) = \varepsilon_p + E_{pp}, E(P^{T_{2g}}) = \varepsilon_p - E_{pp}$ where $\varepsilon_p$ is the atomic $p$-level, and $E_{pp} = t_{pp\sigma} - t_{pp\pi}$.
For each orbital channel, the $e_g-P^{E_g}$ and $t_{2g}-P^{t_{2g}}$ couplings blocks are
\begin{gather}
H_{e_g-P^{E_g}} = \begin{bmatrix} \varepsilon_{e_g} & \sqrt{3}t_{pd\sigma} \\ \sqrt{3}t_{pd\sigma} &E(P^{E_g}) \end{bmatrix}, \quad H_{t_{2g}-P^{t_{2g}}} = \begin{bmatrix} \varepsilon_{t_{2g}} & 2t_{pd\pi} \\ 2t_{pd\pi} &E(P^{t_{2g}}) \end{bmatrix} .
\end{gather}
The higher-level antibonding $e_g^*$ and $t_{2g}\pi*$ levels are therefore
\begin{gather}
E_{e_g^*} = \frac{\varepsilon_{e_g}+ E(P^{E_g})}{2} + \frac{1}{2}\sqrt{(\varepsilon_{e_g} - E(P^{E_g}))^2  + 12t_{pd\sigma}^2}, \\
E_{t_{2g}\pi*} = \frac{\varepsilon_{t_{2g}}+ E(P^{T_{2g}})}{2} + \frac{1}{2}\sqrt{(\varepsilon_{t_{2g}} - E(P^{T_{2g}}))^2  + 16t_{pd\pi}^2}.
\end{gather}
The corresponding antibonding states are
\begin{gather}
|e_{g,\alpha}^*\rangle_0 \propto (E_{e_g^*} - E(P^{E_g})) |e_{g,\alpha}\rangle + (\sqrt{3}t_{pd\sigma}) |P^{E_g}_{\alpha}\rangle, \\
|t_{2g,\alpha}*\rangle_0 \propto (E_{t_{2g}*} - E(P^{T_{2g}})) |t_{2g,\alpha}\rangle + (2t_{pd\pi}) |P^{T_{2g}}_{\alpha}\rangle ,
\end{gather}
where $\alpha = X^2-Y^2, 3Z^2-R^2$ for $e_g$ and $\alpha = XY, YZ, XZ$ for $t_{2g}$. 

\subsection{Spin-orbital tail of the molecular wavefunction}
With inclusion of the ligand SOC, the Wannier function acquire a spin-orbital tail from the neighboring ligand $p$ orbital. Here, we focus on the unperturbed $|e_{g,\alpha}^*\rangle_0$ states and treat $H_{SOC}^p = \lambda_p \vec{L}\cdot \vec{S}$ as a perturbation, we obtain the perturbed states
\begin{gather}
|e_{g\alpha}*, \sigma \rangle =u|e_{g\alpha}, \sigma \rangle + v|P_{\alpha}^{E_g}, \sigma \rangle + |\delta e_{g\alpha}*, \sigma \rangle,
\end{gather}
where
\begin{gather}
u = \frac{\Delta_* - E_{pp}}{\sqrt{(\Delta_*-E_{pp})^2+ 3t_{pd\sigma}^2}}, \quad v = \frac{\sqrt{3}t_{pd\sigma}}{\sqrt{(\Delta_*-E_{pp})^2+ 3t_{pd\sigma}^2}} .
\end{gather}
To first-order in perturbation,
\begin{gather}
|\delta e_{g\alpha}*, \sigma \rangle = \sum_{m} |m\rangle \frac{\langle n| H_{SOC}^p|e_{g\alpha,\sigma}*\rangle_0}{E_{eg*}- E_m} = v \sum_{n\in p} \frac{\langle n| \lambda_p \vec{L}\cdot \vec{S}| P_{\alpha}^{E_g} \rangle}{E_{eg*} - E_n}.
\end{gather}
For convenience, we further define the following ligand orbitals,
\begin{gather}
\begin{aligned}
|Q_{XY},\sigma\rangle &= \frac{1}{2} \big[ |X_-\rangle |p_{Y}, \sigma \rangle - |Y_-\rangle |p_{X}, \sigma \rangle \big], \\
|Q_{YZ},\sigma\rangle &=\frac{1}{2} \big[|Y_-\rangle |p_{Z}, \sigma\rangle - |Z_-\rangle |p_{Y}, \sigma \rangle \big], \\
|Q_{XZ},\sigma\rangle &= \frac{1}{2} \big[ |X_-\rangle |p_{Z}, \sigma\rangle - |Z_-\rangle |p_{X}, \sigma\rangle \big].
\end{aligned}
\end{gather}
with energy $E(Q) = \varepsilon_p + E_{pp}$.
The first-order SOC-induced corrections $|\delta e_{g\alpha}^*, \sigma\rangle$ are given by
\begin{gather}
\begin{aligned}
|\delta d_{X^2-Y^2}*, \sigma\rangle = &\bigg\{ \frac{\lambda_p}{4(\Delta_*+E_{pp})}\bigg[ -\frac{i}{\sqrt{2}}\bigg(P_{YZ}-P_{XZ}\bigg)\sigma_x - \frac{i}{\sqrt{6}}\bigg(4P_{XY}+P_{YZ}+P_{XZ} \bigg)\sigma_y + \frac{i}{\sqrt{3}}\bigg(2P_{XY} - P_{YZ} - P_{XZ} \bigg)\sigma_z \bigg]\\
&+ \frac{\lambda_p}{4(\Delta_*-E_{pp})}\bigg[ -\frac{i}{\sqrt{2}} \bigg(|Q_{YZ}\rangle-|Q_{XZ}\rangle\bigg) \sigma_x -\frac{i}{\sqrt{6}}\bigg(|Q_{YZ}\rangle+ |Q_{XZ}\rangle \bigg)\sigma_y - \frac{i}{\sqrt{3}} \bigg(|Q_{YZ}\rangle+ |Q_{XZ}\rangle \bigg)\sigma_z \bigg] \bigg\} |\sigma\rangle
\end{aligned}
\end{gather}
and
\begin{gather}
\begin{aligned}
|\delta d_{3Z^2-R^2}, \sigma\rangle= &\bigg\{ \frac{\lambda_p}{2\sqrt{12}} \frac{1}{(\Delta_*+E_{pp})} \bigg[ -\frac{3i}{\sqrt{2}} \bigg(|P_{YZ}\rangle + |P_{XZ}\rangle \bigg)\sigma_x + \frac{3i}{\sqrt{6}}\bigg(|P_{XZ}\rangle - |P_{YZ}\rangle \bigg)\sigma_y+ \frac{3i}{\sqrt{3}}\bigg(|P_{XZ}\rangle -|P_{YZ}\rangle \bigg)\sigma_z \bigg] \\
&+ \frac{\lambda_p}{2\sqrt{12}(\Delta_*-E_{pp})} \bigg[ \frac{i}{\sqrt{2}}\bigg(|Q_{YZ}\rangle + |Q_{XZ}\rangle \bigg)\sigma_x + \frac{i}{\sqrt{6}}\bigg(-4|Q_{XY}\rangle + |Q_{YZ}\rangle - |Q_{XZ}\rangle \bigg)\sigma_y \\
&\hspace{10em} +\frac{i}{\sqrt{3}} \bigg(-2|Q_{XY}\rangle + |Q_{YZ}\rangle - |Q_{XZ}\rangle \bigg) \sigma_z \bigg] \bigg\} |\sigma\rangle .
\end{aligned}
\end{gather}
The effective orbitals $|a, \sigma\rangle$ and $|b, \sigma \rangle$ are defined as
\begin{gather}\label{expression of the effective two orbitals}
\begin{aligned}
|a, \sigma \rangle &= u|d_{3Z^2-R^2}, \sigma \rangle + v|P_{3Z^2-R^2}, \sigma\rangle + |\delta d_{3Z^2-R^2}*, \sigma\rangle, \\
|b, \sigma \rangle &= u|d_{X^2-Y^2}, \sigma \rangle + v|P_{X^2-Y^2}, \sigma \rangle + |\delta d_{X^2-Y^2}*, \sigma \rangle.
\end{aligned}
\end{gather}
Using the DFT-derived parameters of the $dp$ model listed in Table.~\ref{dp_model DFT parameters}, we compute the weights $|u|^2$ and $|v|^2$ of the effective orbitals in $\mathrm{NiX_2}$ materials, as summarized in Table.~\ref{Table_dp_weight}. These quantities correspond to Ni-$d$ and ligand-$p$ weights, respectively, in the antibonding $e_g^*$ states.
\begin{table}[H]
\centering
\begin{tabular}{|l|l|l|}
\hline
 ~\text{Weight}~ & ~$|u|^2$~ &~$|v|^2$~  \\ \hhline{|=|=|=|}
~$\text{NiCl}_2$~ &~0.721 ~      &~0.279 ~  \\ \hline
 ~$\text{NiBr}_2$~ &~0.689  ~     &~0.311 ~  \\ \hline
 ~$\text{NiI}_2$~ &~0.569  ~     &~0.431 ~  \\ \hline
\end{tabular}
\caption{The Ni-$d$ weight ($|u|^2$) and ligand-$p$ weight ($|v|^2$) of the antibonding $e_g^*$ states, computed from the $\mathrm{NiX_6}$ cluster, and using DFT-derived $dp$-model parameters.}
\label{Table_dp_weight}
\end{table}

\subsection{Electric polarization in the two-orbital model: beyond generalized Katsura-Nagaosa-Balatsky framework}
We follow the discussion in the the Supplementary Material of \cite{solovyev2021magnetically}, where the polarization is formulated within the framework of superexchange theory. In their formulation, the Wannier function acquires tails on the neighboring sites to first-order in $\hat{t}/U$, giving rise to the electric polarization. Here, we extend the calculations to our two-orbital $e_g$ model. \\

Let $|\chi_i\rangle$ be the spin-polarized states at the site $i$ and $|\chi_i^\perp\rangle$ be its complement, which are given by
\begin{gather}
|\chi_i\rangle = e^{-i\phi_i}\cos \frac{\theta_i}{2} |\ua\rangle + \sin \frac{\theta_i}{2} |\da\rangle , \quad
|\chi_i^\perp \rangle = - \sin \frac{\theta_i}{2}|\ua\rangle + e^{i\phi_i} \cos \frac{\theta_i}{2}|\da \rangle .
\end{gather}
We start from the product state of the local $S = 1$ ground states,
\begin{gather}\label{spin-polarized S=1 ground state}
|g\rangle = \prod_i |g_i\rangle = \prod_i |a, \chi_i\rangle |b, \chi_i\rangle,
\end{gather}
where the spins in the two orbitals $(a,b)$ are aligned by Hund's coupling.
In the Mott limit, the polarization can be calculated by treating the hopping $\hat{T}$ as a perturbation. The resulting polarization is given by
\begin{gather}
\Vector{P} = (-e) P \left( \hat{\Vector{r}} Q \frac{1}{E_0 - H_0} Q\hat{T}  + \hat{T} Q\frac{1}{E_0 - H_0} Q \hat{\Vector{r}} \right)P,
\end{gather}
where $\hat{\Vector{r}}$ denote the position operator, and
\begin{gather}
P = \prod_{i} |g_i\rangle \langle g_i | , \quad Q \equiv 1-P ,
\end{gather}
are the projection operators onto the spin-polarized ground state and its complement, respectively. Restricted to the $(ij)$-bond, the polarization reduces to
\begin{gather}\label{Superexchange expression for bond polarization}
\Vector{P}_{ij} = e \frac{1}{(U+J_H)} \langle g_i g_j| \hat{\Vector{r}}_{(ij)} \hat{T}_{(ij)} + \hat{T}_{(ij)} \hat{\Vector{r}}_{(ij)} |g_i g_j\rangle ,
\end{gather}
where $\hat{T}_{(ij)}$ is the hopping Hamiltonian on the bond $(ij)$, and the bond position operator is
\begin{gather}
\hat{\Vector{r}}_{(ij)} = \Psi_i^\dag \Vector{r}_{ij} \Psi_j + \Psi_j^\dag \Vector{r}_{ij}^\dag \Psi_i .
\end{gather}
where  $\Psi_i^\dag = [a_{i\ua}^\dag, a_{i\da}^\dag, b_{i\ua}^\dag, b_{i\da}^\dag]$ is the the four-component spinor at site $i$.
Inversion symmetry and Hermiticity of $\hat{\Vector{r}}_{(ij)}$ requires the matrices to be anti-Hermitian, i.e, $\Vector{r}_{ij}^\dag = -\Vector{r}_{ij}$. Focusing on the $\delta_1$-bond, the $C_{2x}$ bond symmetry further constrains the allowed terms to be
\begin{align}
r_{ij}^x &= R^x_{0x}i\tau_0 \sigma_x+  R^x_{zx}i\tau_z \sigma_x+ R^x_{xy} i\tau_x \sigma_y+ R^x_{xz} i\tau_x \sigma_z = i\begin{bmatrix} R_{ax}^x &0 \\ 0 &R_{bx}^x \end{bmatrix} \sigma_x + R^x_{xy} i\tau_x \sigma_y+ R^x_{xz} i\tau_x \sigma_z, \label{symmetry-allowed position operator matrix 1} \\
r_{ij}^{\alpha} &= R_{y0}^{\alpha} i\tau_y \sigma_0 + R_{0y}^{\alpha}i\tau_0 \sigma_y+ R_{0z}^{\alpha} i\tau_0 \sigma_z \notag + R_{zy}^{\alpha} i\tau_z \sigma_y+ R_{zz}^{\alpha} i\tau_z \sigma_z+ R_{xx}^{\alpha} i\tau_x \sigma_x\\
&= R_{y0}^{\alpha} i\tau_y \sigma_0 + i\begin{bmatrix} R_{ay}^{\alpha} &0 \\ 0 &R_{by}^{\alpha} \end{bmatrix} \sigma_y + i\begin{bmatrix} R_{az}^{\alpha} &0 \\ 0 &R_{bz}^{\alpha} \end{bmatrix} \sigma_z + R_{xx}^{\alpha} i\tau_x \sigma_x \quad (\alpha = y,z), \label{symmetry-allowed position operator matrix 2}
\end{align}
where $\tau_{0,x,y,z}$ and $\sigma_{0,x,y,z}$ are Pauli matrices in the orbital and spin spaces, respectively, and the coefficients $R_{\alpha \beta}^{\gamma}$ are real numbers. We recall from the main text, the bond hopping matrix along the $\delta_1$-bond is given by
\begin{gather}
T^\pdg_{ij} = \begin{bmatrix} \!- t_{1a} &0 \\ 0 &\! - t_{1b}\end{bmatrix} \sigma_0 +
\lambda_{1z} \tau_y \sigma_z- \lambda_{1y} \tau_y \sigma_y .
\end{gather}
Then Eq.~\eqref{Superexchange expression for bond polarization} is determined by expressions of the form 
\begin{gather}\label{derivation eq1 polarization}
\langle g_i g_j| \Psi_i^\dag A \Psi_j \Psi_j^\dag B \Psi_j|g_i g_j \rangle,
\end{gather}
where $A,B$ are $4\times 4$ matrices from $\hat{r}_{(ij)}$ and $\hat{T}_{(ij)}$. With the spin-polarized $S=1$ ground state considered in Eq.~\eqref{spin-polarized S=1 ground state}, this simplify to
\begin{gather}
\langle g_i g_j| \Psi_i^\dag A \Psi_j \Psi_j^\dag B \Psi_j|g_i g_j \rangle = \langle g_i| \Psi_i^\dag (A \big(\mathbb{I} - P_{j}^{\text{occ}}\big) B) \Psi_i |g_i\rangle,
\end{gather}
where
\begin{gather}
(\mathbb{I} - P_j^{\text{occ}}) = \sum_{k=a,b} |k,\chi_j^\perp\rangle \langle k,\chi_j^\perp| .
\end{gather}
Then Eq.~\eqref{derivation eq1 polarization} becomes
\begin{gather}
\langle g_i g_j| \Psi_i^\dag A \Psi_j \Psi_j^\dag B \Psi_i|g_i g_j \rangle = \sum_{k=a,b}\langle a, \chi_{i}| A|k,\chi_j^\perp\rangle \langle k, \chi_j^\perp|B|a, \chi_{i} \rangle + \langle b, \chi_{i}| A|k,\chi_j^\perp\rangle \langle k, \chi_j^\perp|B|b, \chi_{i} \rangle.
\end{gather}
With $A$ and $B$ of the form $\tau_m \sigma_n$, we note the following matrix elements:
\begin{gather}
\langle \alpha, \chi_{i}| \tau_{m} \sigma_n|\beta, \chi_{j}^\perp \rangle = \langle \alpha| \tau_m|\beta \rangle \langle\chi_i|\sigma_n|\chi_j^\perp \rangle,
\end{gather}
where
\begin{gather}
\langle \chi_i|\sigma_0|\chi_j^\perp\rangle = -\cos \frac{\theta_i}{2} \sin \frac{\theta_j}{2} e^{i\phi_i} + \sin \frac{\theta_i}{2} \cos \frac{\theta_j}{2} e^{-i\phi_j}, \\
\langle \chi_i|\sigma_x|\chi_j^\perp \rangle = -\sin\frac{\theta_i}{2}\sin \frac{\theta_j}{2} + e^{i(\phi_i+\phi_j)} \cos \frac{\theta_i}{2} \cos \frac{\theta_j}{2}, \\
\langle \chi_i|\sigma_y|\chi_j^\perp \rangle = i\left( \cos \frac{\theta_i}{2} \sin \frac{\theta_j}{2} e^{i\phi_i} - \sin \frac{\theta_i}{2} \cos \frac{\theta_j}{2} e^{i\phi_j} \right), \\
\langle \chi_i|\sigma_z|\chi_j^\perp \rangle = -\cos \frac{\theta_i}{2} \sin \frac{\theta_j}{2} e^{i\phi_i} - \sin \frac{\theta_i}{2} \cos \frac{\theta_j}{2} e^{-i\phi_j} .
\end{gather}
This gives
\begin{gather}
\langle \chi_i|\sigma_0|\chi_j^\perp \rangle \langle \chi_j^\perp| \sigma_0|\chi_i\rangle = \frac{1}{2}\big(1- \vec{e}_i \cdot \vec{e}_j\big), \\
\langle \chi_i|\sigma_m|\chi_j^\perp\rangle \langle \chi_j^\perp|\sigma_m|\chi_i\rangle  = \frac{1}{2}\big(1+\vec{e}_i\cdot \vec{e}_j\big) - e_i^m e_j^m, \\
\langle \chi_i| \sigma_m|\chi_j^\perp \langle \chi_j^\perp|\sigma_n|\chi_i\rangle + h.c. = -\big(e_i^m e_j^n + e_i^n e_j^m\big) \quad (m \neq n) , \\
i\langle \chi_i|\sigma_l|\chi_j^\perp\rangle \langle \chi_j^\perp|\sigma_0|\chi_i\rangle + h.c. = - \big(\vec{e}_i \times \vec{e}_j\big)_l.\label{antisymmetric spin component}
\end{gather}
As discussed in the main text, with the local $S = 1$ ground state and centrosymmetric bond, only the spin-antisymmetric component Eq.~\eqref{antisymmetric spin component} survives, and the polarization is simplify to
\begin{gather}
\Vector{P}_{ij} = e \frac{1}{(U+J_H)} \langle g_i g_j| \hat{\Vector{r}}_{(ij)}^{\text{SD}} \hat{T}_{(ij)}^{\text{SI}}+ \hat{\Vector{r}}_{(ij)}^{\text{SI}} \hat{T}_{(ij)}^{\text{SD}} + h.c. |g_i g_j\rangle ,
\end{gather}
where SI and SD denote the spin-independent and spin-dependent components, respectively. This has the same structure as in the one-orbital case discussed in Ref.~\cite{solovyev2021magnetically}, where the polarization arises from the interference between SI and SD terms. In the present two-orbital model, however, both $\hat{\Vector{r}}_{ij}$ and $\hat{T}_{(ij)}$ contain a richer set of symmetry-allowed SI and SD contributions. Using these results, we obtain the $D_{ij}$ matrix for the 1NN $\delta_1$-bond given in the main text.

\subsection{Derivation of the position operator coefficients}
In this section, we first identify the nonzero polarization integrals in the full $d$-$p$ basis, and then derive the coefficients $R_{\alpha \beta}^{\gamma}$ in Eq.~\eqref{symmetry-allowed position operator matrix 1}-\eqref{symmetry-allowed position operator matrix 2} for the effective two orbital model. \\

Focusing on the $\delta_1$-bond, the $C_{2x}$ act as $(X,Y,Z) \to (-Y, -X, -Z), \quad (x,y,z)\to (x,-y,-z)$, under which the orbitals have the following $C_{2x}$ parities:
\begin{gather}
\begin{aligned}
\begin{cases}
\text{odd: } |d_{X^2-Y^2}\rangle, \quad |P_{X^2-Y^2}\rangle, \quad |P_{YZ}\rangle - |P_{XZ}\rangle, \quad |Q_{XY}\rangle, \quad |Q_{YZ}\rangle - |Q_{XZ}\rangle \\
\text{even: } |d_{3Z^2-R^2}\rangle, \quad |P_{3Z^2-R^2}\rangle, \quad |P_{YZ}\rangle + |P_{XZ}\rangle, \quad |P_{XY}\rangle, \quad |Q_{YZ}\rangle + |Q_{XZ}\rangle
\end{cases}.
\end{aligned}
\end{gather}
The $C_{2x}$ symmetry then constrains the allowed polarization integrals $\langle M_i|\Vector{r}|N_j\rangle$. Since $x$ is even under $C_{2x}$, and $y,z$ are odd under $C_{2x}$, the following matrix elements are nonzero:
\begin{gather}
\begin{aligned}
<M_i|x|N_j\rangle: \quad 
&|M_i\rangle \in \big\{|d_{X^2-Y^2}\rangle, |P_{X^2-Y^2}\rangle \big\}, \quad |N_j\rangle \in \big\{ C_{2x}\text{-odd orbitals} \big\} \\
&|M_i\rangle \in \big\{ |d_{3Z^2-R^2}\rangle, |P_{3Z^2-R^2}\rangle \big\}, \quad |N_j\rangle \in \big\{ C_{2x}\text{-even orbitals} \big\}
\end{aligned}, \\[5pt]
\begin{aligned}
<M_i|y/z|N_j\rangle: \quad 
&|M_i\rangle \in \big\{|d_{X^2-Y^2}\rangle, |P_{X^2-Y^2}\rangle \big\}, \quad |N_j\rangle \in \big\{ C_{2x}\text{-even orbitals} \big\} \\
&|M_i\rangle \in \big\{|d_{3Z^2-R^2}\rangle, |P_{3Z^2-R^2}\rangle \big\}, \quad |N_j\rangle \in \big\{ C_{2x}\text{-odd orbitals} \big\}
\end{aligned}.
\end{gather}

Using the spin-orbital tails of the effective two orbitals derived in Eq.~\eqref{expression of the effective two orbitals}, the matrix elements $R_{\alpha \beta}^{\gamma}$ in Eq.~\eqref{symmetry-allowed position operator matrix 1}-\eqref{symmetry-allowed position operator matrix 2} can then be expressed in terms of these polarization integrals.
For example, the spin-independent components are given by
\begin{gather}
\begin{aligned}
R_{y0}^{y(z)} = &u^2\langle d_{3Z^2-R^2}|y(z)|d_{X^2-Y^2}\rangle + v^2 \langle P_{3Z^2-R^2}|y(z)|P_{X^2-Y^2}\rangle \\
&+ uv \langle d_{3Z^2-R^2}|y(z)|P_{X^2-Y^2}\rangle + uv \langle P_{3Z^2-R^2}|y(z)|d_{X^2-Y^2}\rangle.
\end{aligned}
\end{gather}
An example of a spin-dependent component is
\begin{gather}
\begin{aligned}
R^x_{ax} = &uv\frac{\lambda_p}{2\sqrt{12}}\frac{1}{(\Delta_*+E_{pp})} \sqrt{2} \bigg( -3 \langle d_{3Z^2-R^2},i|x| \big(|P_{YZ}, j\rangle + |P_{XZ}, j\rangle \big) \bigg) \\
&+ uv \frac{\lambda_p}{2\sqrt{12}(\Delta_*-E_{pp})} \sqrt{2} \bigg( \langle d_{3Z^2-R^2},i|x| \big(|Q_{YZ}, j\rangle + |Q_{XZ}, j\rangle \big) \bigg) \\
&+v^2\frac{\lambda_p}{2\sqrt{12}}\frac{1}{(\Delta_*+E_{pp})} \sqrt{2} \bigg( -3 \langle P_{3Z^2-R^2},i|x| \big(|P_{YZ}, j\rangle + |P_{XZ}, j\rangle \big) \bigg) \\
&+ v^2 \frac{\lambda_p}{2\sqrt{12}(\Delta_*-E_{pp})} \sqrt{2} \bigg( \langle P_{3Z^2-R^2},i|x| \big(|Q_{YZ}, j\rangle + |Q_{XZ}, j\rangle \big) \bigg) .
\end{aligned}
\end{gather}
Similar expressions can be obtained for other spin-dependent components. We estimate the corresponding values of $R_{\alpha \beta}^\gamma$ for $\mathrm{NiI_2}$ using hydrogenlike atomic orbitals, taking the bond lengths to be $d(M-M) = d(X-X) = 3.98\AA$ and $d(M-X) = 2.81\AA$. We use Clementi-Raimondi effective charges $Z^{eff}_{\mathrm{Ni, 3d}} = 12.53$ and $Z^{eff}_{\mathrm{I, 5p}} = 11.61$. For the 1NN bond, in units of Bohr radius $(a_0)$, we find
\begin{gather}
\begin{aligned}
&\text{1NN}: R_{ax}^x = R_{bx}^x = R_{xx}^y = R_{xx}^z = 0, \\
&R_{ay}^y = 0.0266 a_0, \quad R_{by}^y = -0.0411a_0, \quad R_{ay}^z= 0.0325 a_0, \quad R_{by}^z = -0.0622a_0, \\
&R_{az}^y = -0.0127a_0, \quad R_{bz}^y = 0.0333 a_0, \quad R_{az}^z = -0.0266a_0, \quad R_{bz}^z = 0.0405a_0, \\
&R_{y0}^y = 0.02085 a_0, \quad R_{y0}^z = 0.02929a_0.
\end{aligned}
\end{gather}
And for the 3NN bond, we find
\begin{gather}
\begin{aligned}
&\text{3NN}: R_{ax}^x = R_{bx}^x = R_{xx}^y = R_{xx}^z = 0, \\
&R_{ay}^y = -0.0023 a_0, \quad R_{by}^y = 0.00395a_0, \quad R_{ay}^z= -0.0031 a_0, \quad R_{by}^z = 0.00364 a_0, \\
&R_{az}^y = 0.0016a_0, \quad R_{bz}^y = -0.0008 a_0, \quad R_{az}^z = 0.0025a_0, \quad R_{bz}^z = -0.0043a_0, \\
&R_{y0}^y = 0.04945 a_0, \quad R_{y0}^z = 0.06933 a_0.
\end{aligned}
\end{gather}
We summarize some of the key numerical results from this section in the main text.

\end{document}